\documentclass[fleqn,usenatbib]{mnras}

\usepackage[T1]{fontenc}

\DeclareRobustCommand{\VAN}[3]{#2}
\let\VANthebibliography\thebibliography
\def\thebibliography{\DeclareRobustCommand{\VAN}[3]{##3}\VANthebibliography}

\usepackage{graphicx}	
\usepackage{amsmath}	
\usepackage{siunitx}    
\usepackage{subfig}     
\usepackage{newtxtext,newtxmath}

\newcommand{\heii}{He\,\textsc{ii}}
\newcommand{\lya}{Ly$\alpha$}
\newcommand{\civ}{C\,\textsc{iv}}
\newcommand{\ciii}{C\,\textsc{iii}]}
\newcommand{\oii}{[O\,\textsc{ii}]}
\newcommand{\oiii}{[O\,\textsc{iii}]}
\newcommand{\hb}{H$\beta$}
\newcommand{\ohb}{[O\,\textsc{iii}]\,+\,H$\beta$}

\newcommand{\sfr}{$M_\odot$\,yr$^{-1}$}

\newcommand{\fesc}{$f_\textrm{esc}$}
\newcommand{\fescrel}{$f_{\textrm{esc}}^{\textrm{rel}}$}

\title[LyC leakage at $z\sim3-3.5$]{No strong dependence of Lyman continuum leakage on physical properties of star-forming galaxies at $\mathbf{3.1 \lesssim z \lesssim 3.5}$}

\author[A. Saxena et al.]{A. Saxena,$^{1}$\thanks{E-mail: aayush.saxena@ucl.ac.uk}
L. Pentericci,$^{2}$
R. S. Ellis,$^{1}$
L. Guaita,$^{3}$
A. Calabr\`{o},$^{2}$
D. Schaerer,$^{4,5}$
E. Vanzella,$^{6}$
\newauthor R. Amor\'{i}n,$^{7,8}$
M. Bolzonella,$^{9}$
M. Castellano,$^{2}$
F. Fontanot,$^{10,11}$
N. P. Hathi,$^{12}$
P. Hibon,$^{13}$
M. Llerena,$^{8}$
\newauthor F. Mannucci,$^{14}$
A. Saldana-Lopez,$^{4}$
M. Talia,$^{15}$
G. Zamorani$^{9}$\\ \\
$^{1}$Department of Physics and Astronomy, University College London, Gower Street, London WC1E 6BT, UK\\	
$^{2}$INAF -- Osservatorio Astronomico di Roma, via Frascati 33, 00078, Monteporzio Catone, Italy \\
$^3$Departamento de Ciencias Fisicas, Universidad Andres Bello, Fernandez Concha 700, Las Condes, Santiago, Chile \\
$^{4}$Observatoire de Gen\`eve, D\'epartement d’Astronomie, Universit\'e de Gen\`eve, 51 Chemin Pegasi, 1290 Versoix, Switzerland \\
$^{5}$CNRS, IRAP, 14 avenue E. Belin, 31400 Toulouse, France \\
$^{6}$INAF – Osservatorio Astronomico di Bologna, Via P. Gobetti 93/3, 40129 Bologna, Italy \\
$^{7}$Instituto de Investigaci\'{o}n Multidisciplinar en Ciencia y Tecnolog\'{i}a, Universidad de La Serena, Ra\'{u}l Bitr\'{a}n, 1305 La Serena, Chile \\
$^{8}$Departamento de F\'{i}sica y Astronom\'{i}a, Universidad de La Serena, Av. Juan Cisternas 1200 Norte, La Serena, Chile \\
$^{9}$INAF - Osservatorio di Astrofisica e Scienza dello Spazio di Bologna, via Piero Gobetti 93/3, 40129 Bologna, Italy \\
$^{10}$INAF –- Astronomical Observatory of Trieste, Via G.B. Tiepolo 11, 34143 Trieste, Italy \\
$^{11}$IFPU –- Institute for Fundamental Physics of the Universe, Via Beirut 2, 34151 Trieste, Italy \\
$^{12}$Space Telescope Science Institute, 3700 San Martin Drive, Baltimore, MD 21218, USA \\
$^{13}$European Southern Observatory (ESO), Vitacura, Chile \\
$^{14}$INAF – Osservatorio Astrofisico di Arcetri, Largo E. Fermi 5, 50125 Firenze, Italy \\
$^{15}$University of Bologna -- Department of Physics and Astronomy ``Augusto Righi'' (DIFA), Via Gobetti 93/2, 40129 Bologna, Italy
}

\date{Accepted 2021 December 19. Received 2021 November 23; in original form 2021 September 09}

\pubyear{2021}

\begin{document}
\label{firstpage}
\pagerange{\pageref{firstpage}--\pageref{lastpage}}
\maketitle

\begin{abstract}
We present Lyman continuum (LyC) radiation escape fraction (\fesc) measurements for 183 spectroscopically confirmed star-forming galaxies in the redshift range $3.11 < z < 3.53$ in the \textit{Chandra} Deep Field South. We use ground-based imaging to measure \fesc, and use ground- and space-based photometry to derive galaxy physical properties using spectral energy distribution (SED) fitting. We additionally derive \ohb\ equivalent widths (that fall in the observed $K$ band) by including nebular emission in SED fitting. After removing foreground contaminants, we report the discovery of 11 new candidate LyC leakers at $\gtrsim2\sigma$ level, with \fesc\ in the range $0.14-0.85$. From non-detections, we place $1\sigma$ upper limits of \fesc\,$<0.12$, where the Lyman-break selected galaxies have \fesc\,$<0.11$ and `blindly' discovered galaxies with no prior photometric selection have \fesc\,$<0.13$. We find a slightly higher $1\sigma$ limit of \fesc\,$<0.20$ from extreme emission line galaxies with rest-frame \ohb\ equivalent widths $>300$\,\AA. For candidate LyC leakers, we find a weak negative correlation between \fesc\ and galaxy stellar masses, no correlation between \fesc\ and specific star-formation rates (sSFRs) and a positive correlation between \fesc\ and EW$_0$(\ohb). The weak/no correlations between stellar mass and sSFRs may be explained by misaligned viewing angles and/or non-coincident timescales of starburst activity and periods of high \fesc. Alternatively, escaping radiation may predominantly occur in highly localised star-forming regions, or \fesc\ measurements may be impacted by stochasticity of the intervening neutral medium, obscuring any global trends with galaxy properties. These hypotheses have important consequences for models of reionisation.
\end{abstract}

\begin{keywords}
cosmology: dark ages, reionisation, first stars -- cosmology: early Universe -- galaxies: high-redshift -- galaxies: evolution
\end{keywords}



\section{Introduction}
\label{sec:introduction}
Constraining when and how the intergalactic medium (IGM) in the Universe made a phase transition from neutral to completely ionised, i.e. the epoch of reionisation (EoR) is an exciting challenge in observational cosmology. An important period in the evolutionary history of our Universe, EoR was likely triggered by UV photon production by the first stars, galaxies and supermassive black holes in the Universe \citep{bro11} and observations suggest that the reionisation of the Universe unfolded over the redshift range $6\lesssim z \lesssim 10$ \citep[e.g.][]{planck_eor}. To understand how reionisation progressed, knowledge of the intrinsic ionising spectra of sources in the early Universe is needed, in addition to the fraction of Hydrogen-ionising Lyman continuum (LyC; $\lambda_0 < 912$\,\AA) photons that actually escape into the neutral IGM driving reionisation \citep[see][for a recent review]{day18}. The rapidly increasing neutrality of the IGM at high redshifts \citep[e.g.][]{ino14} makes it impossible to directly observe LyC photons from reionisation era galaxies even with facilities like the \emph{James Webb Space Telescope (JWST)} and therefore, the only possibility to obtain constraints on the escape fraction of LyC photons (\fesc) is by studying lower redshift analogues of reionisation era galaxies, where escaping LyC photons are not completely absorbed by the comparatively less neutral IGM.

Assuming a constant LyC escape fraction for galaxies across the full range of UV luminosities or stellar masses, various works (\citealt{rob13, rob15}, but see also \citealt{nai20}) have determined that \fesc\ $\approx10-20\%$ is required to both reionise the Universe by $z\sim6$ and match the Thomson optical depth of electron scattering observed in the cosmic microwave background \citep{planck_eor}. High-resolution zoom-in simulations of reionisation era galaxies have suggested that \fesc\ could be higher in lower mass systems, as the energy and momentum ejected by supernova explosions are able to clear out channels of neutral gas more efficiently in lower mass environments, facilitating LyC photon escape \citep{wis14, tre17}. Other simulations, however, have suggested a non-monotonic dependence of \fesc\ on stellar/halo mass, where \fesc\ is lower at the low mass end ($M_\star < 10^8\,M_\odot$) due to inefficient star formation, peaks at $M_\star \sim 10^8\,M_\odot$ and decreases again at higher masses due to dust attenuation \citep[e.g.][]{ma20}. 

Alternatively, a strong dependence of \fesc\ on the star-formation rate surface density may imply that higher mass galaxies ($>10^8~M_\odot$) at $z>6$ may contribute $>80\%$ of the reionisation photon budget and also explain the rapid conclusion of reionisation \citep{nai20}. Active galactic nuclei (AGN) at $z>6$, with a considerably higher \fesc\ on average compared to star-forming galaxies \citep{cri16, gra18, rom19}, may also contribute to the observed rapid evolution of the IGM neutral fraction at $z>6$ \citep{fon12, gia15, tre18, fin19, day20}, and contributions to the reionisation budget may be expected from low-metallicity X-ray binaries binary populations in high redshift galaxies \citep{mad17, sch19, sax21}. 

However, no conclusive correlations between \fesc\ and stellar masses or star-formation rates have yet been identified observationally. LyC measurements presented by \citet{izo21} for a sample of very low mass ($M_\star < 10^8\,M_\odot$) galaxies at $0.30<z<0.45$ showed that \fesc\ does not strongly depend on the stellar mass over a broad mass range of $10^7 - 10^{10}\,M_\odot$. Radiative transfer modelling has suggested that density-bounded H\,\textsc{ii} regions within star-forming galaxies, traced by high \oiii/\oii\ ratios (O32; \citealt{zac13}), offer ideal environments to achieve high \fesc\ values \citep{jas13, nak14, fai16, izo18a} and although spectroscopic follow-up of candidate LyC leakers at $z\sim3$ has shown that galaxies with high \fesc\ can also show high O32 \citep[e.g.][]{van16b, nak20}, several other physical phenomena that do not require high \fesc\ can also lead to high O32 ratios within galaxies \citep[e.g.][]{sta15, izo17, paa18, jas19}. Therefore, a high O32 does not also necessarily guarantee high \fesc\ \citep{izo18a, nai18, bas19, nak20}.

Further, studies that have attempted to measure \fesc\ from intermediate redshift galaxies using both ground-based \citep[e.g.][]{gra16, gua16, mar17, jap17, nai18} and space-based \citep[e.g.][]{sia15, mos15, rut17, fle19} UV imaging have generally found low \fesc\ values ($<10-20\%$), with no clear differences in \fesc\ measured from photometrically selected Lyman break galaxies (LBGs) or galaxies selected based on strong emission lines. LyC \fesc\ measured from compact Lyman alpha emitting galaxies (LAEs), considered to be direct analogues of reionisation era galaxies, have reported a higher incidence of individual LyC leakers (with \fesc\ $\gtrsim 10\%$; e.g. \citealt{fle19}), but average \fesc\ measured through stacking of non-leakers remains low \citep{mos13, fle19, bia20, smi20} and consistent with that derived for other types of galaxies. The candidate LyC leaking galaxies identified at $z\sim2-4$, with \fesc\ $>20\%$ \citep[e.g.][]{deb16, van16a, van18, sha16, bia17, riv19, fle19, mes20, mar21}, also generally occupy a broad range of stellar masses, star-formation rates and rest-optical nebular line strengths (e.g. \oiii), showing no clear preference or dependence on either of these parameters.

To fully understand the nature of galaxies that drove reionisation, thereby dictating both the spatial and temporal evolution of the neutral fraction of the IGM at $z>6$, it is necessary to uniformly probe the dependence of \fesc\ on galaxy properties using large samples of star-forming galaxies at intermediate redshifts. To do this, a combination of accurate spectroscopic redshifts, deep rest-frame LyC data and multi-band photometry that enables reliable constraints on the spectral energy distributions (SEDs) is needed. This combination is crucial to overcome uncertainties in the measurement of both LyC \fesc\ as well as galaxy properties such as stellar masses, star-formation rates (SFRs), histories, ages, metallicities, dust attenuation, etc., which may also impact the inferred \fesc\ \citep{day18}. Accurate redshifts can also minimise the uncertainties in the IGM transmission along the line-of-sight that can further affect the \fesc\ measurement \citep[e.g.][]{bas21}.  

Therefore, in this study we construct a sample of star-forming galaxies with high quality spectroscopic redshifts at $z=3.11-3.53$ for which deep ground-based imaging probing rest-frame LyC emission is available. Our aim is to uniformly measure LyC \fesc\ and key galaxy properties using SED fitting to explore potential correlations between them. The galaxy sample used in this study probes a range of UV luminosities, stellar masses and star-formation rates, containing galaxies that were photometrically selected for spectroscopic follow-up based on their Lyman break, as well as blindly discovered galaxies from deep imaging-spectroscopy surveys of the sky. Each galaxy benefits from deep \textit{Hubble Space Telescope (HST)} and \textit{Spitzer}/IRAC imaging of their non-ionising wavelengths, which enables highly reliable SED fitting. The redshift range of our sample is further advantageous due to the redshifted \ohb\ emission lines falling in the observed $K$ broadband, enabling the exploration of the connection between \fesc\ and \ohb\ line strengths.

The layout of this paper is as follows. We describe the LyC imaging and the spectroscopic data sets used in this study, along with the sample selection and foreground contamination identification in \S\ref{sec:sample}. In \S\ref{sec:sed} we describe the SED fitting procedure used to derive key physical properties for our galaxies as well as estimate \ohb\ line strengths. In \S\ref{sec:lyc} we describe our LyC \fesc\ measurements and present the results for both individual galaxies as well as \fesc\ measured through stacking. We also present stacked \fesc\ measurements for sub-samples of galaxies, including those for galaxies that show extremely strong \ohb\ emission lines. In \S\ref{sec:results} we explore the dependence of \fesc\ on key galaxy properties such as stellar masses, specific star-formation rates and \ohb\ strengths for both individual \fesc\ measurements and stacks. Finally in \S\ref{sec:discussion} we discuss the implications of our results and discuss mechanisms that likely regulate \fesc.

Throughout this paper, we assume a $\Lambda$CDM cosmology with $\Omega_\textrm{m} = 0.3$ and H$_0 = 67.7$ km s$^{-1}$ Mpc$^{-1}$ taken from \citet{planck}, and use the AB magnitude system \citep{oke83}. All logarithms are in base 10, unless otherwise specified.

\section{Data and sample selection}
\label{sec:sample}
In this work, we focus on the Great Observatories Origins Deep Survey-South (GOODS-S) extragalactic field\footnote{\url{https://www.stsci.edu/science/goods/}}, which benefits from deep photometric and spectroscopic observations of galaxies from ground- and space-based observatories across a wide wavelength range, for example through the Cosmic Assembly Near-infrared Deep Extragalactic Legacy Survey (CANDELS; \citealt{gro11, koe11}). This field also benefits from deep narrow-band \citep{gua16} and broad-band imaging data \citep{non09} probing rest-frame LyC wavelengths at $z\sim3-3.5$, and deep \emph{HST} and \emph{Spitzer} data that enables reliable SED fitting, crucially at wavelengths red-wards of the Balmer break at $z\sim3.5$ (around $4000$\,\AA~rest-frame). We first describe the available LyC filters in this field below, followed by the spectroscopic data used in this study, possible foreground contaminant identification and final galaxy sample creation.

\subsection{LyC filters -- narrow bands and U-band}
\label{sec:filters}
To measure LyC signal, typically measured in the rest-frame wavelength range $860<\lambda<912$\,\AA\ at $z>3$, we use narrow-band imaging data from \citet{gua16} and the publicly available $U$-band imaging data \citep{non09}. We briefly describe the properties of these LyC filters below.

The images from the narrow bands (NB) \emph{NB3727} and \emph{NB396} were first presented in \citet{gua16} and we refer the readers to this paper for full details of the observations, data reduction, calibration and other properties associated with these images. Briefly, the \emph{NB3727} image was obtained using the MOSAIC II instrument at the Cerro Tololo Interamerican Observatory (CTIO), with the exposure time totalling to 36 hours, a total coverage area of \SI{31.5}{\arcmin} $\times$ \SI{31.5}{\arcmin} on the sky and $1\sigma$ detection limit of 27.38\,AB. \emph{NB3727} offers LyC coverage in the wavelength range mentioned above for redshifts $3.11 < z < 3.34$. 

The \emph{NB396} image was obtained using the WFI instrument mounted on the ESO 2.2m Telescope at La Silla, with a total exposure time of 13 hours, covering a subset of the sky area encompassed by the \emph{NB3727} image with a $1\sigma$ detection limit at 27.72\,AB. \emph{NB396} samples LyC signal in the range $3.41 < z < 3.53$.

\begin{figure}
    \centering
    \includegraphics[scale=0.4]{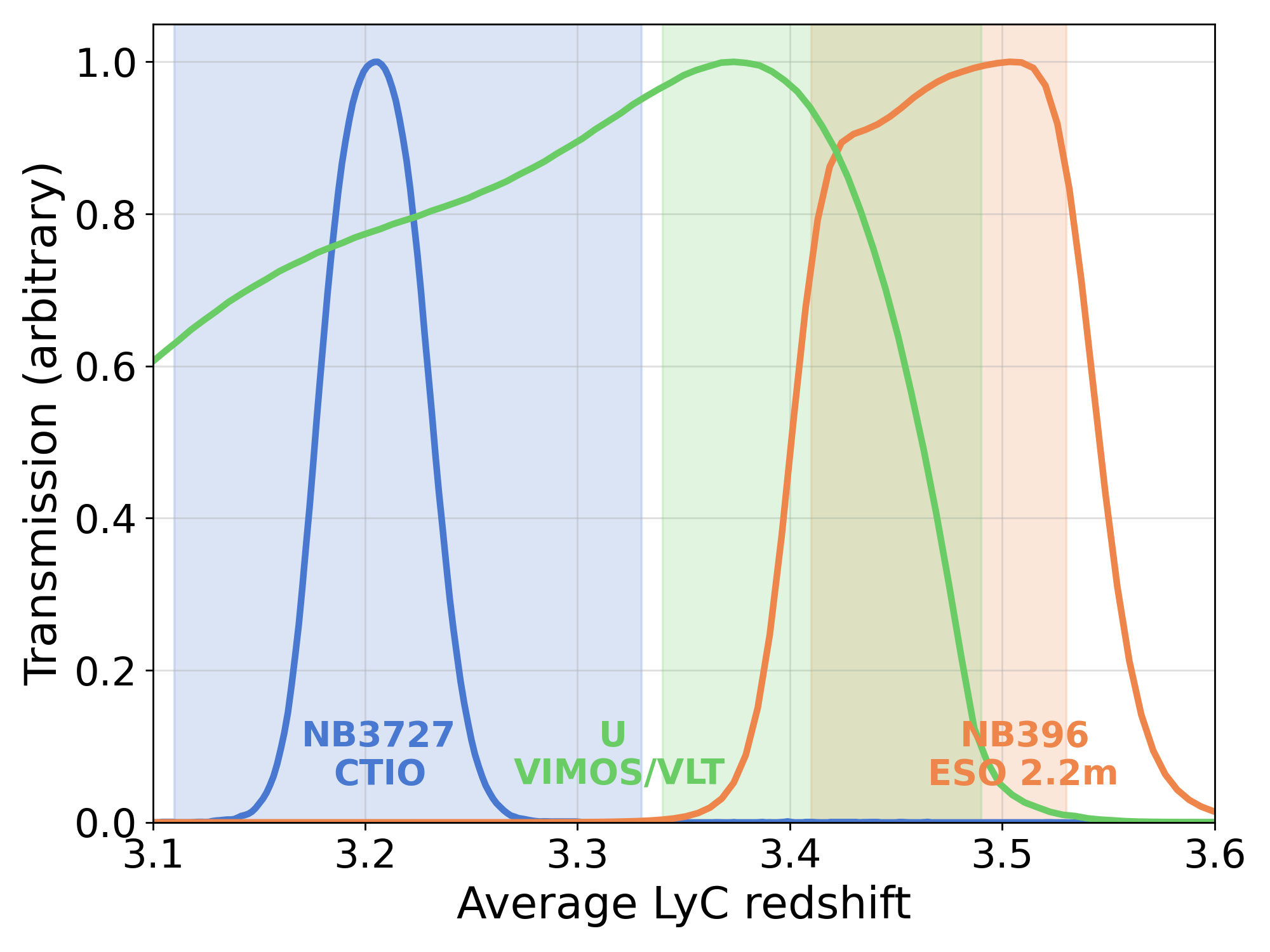}
    \caption{Filter transmission profiles of the \emph{NB3727}, \emph{NB396} and the \emph{U} band filters used to target LyC emission in this study. The x-axis shows the redshift at the rest-frame wavelength of 886\,\AA, which is the mid-point wavelength of the LyC emission sampled between $860 < \lambda < 912$\,\AA\ in this study (shaded regions for each filter). More details about the filter properties and LyC redshift ranges covered are given in \S\ref{sec:filters}.}
    \label{fig:filters}
\end{figure}

We also use the deep $U$-band image in the GOODS-S field taken by \citet{non09} using VIMOS on the \textit{Very Large Telescope (VLT)} covering an area of $\approx630$ arcmin$^2$ and reaching a $1\sigma$ magnitude limit of 29.78\,AB. The $U$-band filter covers rest-frame LyC signal in the redshift range $3.34 < z < 3.49$. Although flux bluewards of the Lyman break is sampled for redshifts above $3.49$, the declining IGM transmissivity would strongly suppress the expected flux at these shorter wavelengths \citep[e.g.][]{ino14} and may bias the measured LyC fluxes. The transmission curves and the redshift range over which the rest-frame LyC flux is sampled by each filter are shown in Figure \ref{fig:filters}.

Therefore, owing to the rest-frame LyC wavelength coverage offered by imaging data, the redshift range probed in this study is $3.11 < z < 3.53$. We now present the spectroscopic data within the GOODS-S/CDFS field used to construct a sample of star-forming galaxies in this redshift range.

\subsection{Spectroscopic data}
\subsubsection{VANDELS}
The majority of spectra used in this study are from VANDELS -- a deep VIMOS survey of the CANDELS UDS and CDFS fields -- which is a recently completed ESO public spectroscopic survey carried out using the VLT. The VANDELS survey covers two well-studied extragalactic fields, the UKIDSS Ultra Deep Survey (UDS) centred around RA = 02:17:38, Dec = $-$05:11:55, and the Chandra Deep Field South (CDFS/GOODS-S) centred around RA = 03:32:30, Dec = $-$27:48:28. As mentioned earlier, we use VANDELS spectra in the CDFS/GOODS-S field in this study. The survey description and target selection can be found in \citet{mcl18}, and detailed data reduction and redshift determination steps can be found in \citet{pen18}.

VANDELS spectroscopic targets were selected based on photometric redshifts (or the presence of Lyman breaks in their broadband photometry) derived from deep ground and space-based multi-band photometry. The parent spectroscopic sample included bright star-forming galaxies ($i \leq 25.0$\,AB) in the range $2.4 \leq z \leq 5.5$ with 99\% of this subsample having specific star-formation rates (sSFR) $>0.6$\,Gyr$^{-1}$, fainter Lyman-break galaxies (LBGs; $i \leq 27.5$\,AB) in the redshift range $3.0 \leq z \leq 7.0$ with sSFR $>0.3$\,Gyr$^{-1}$ and passive galaxies with sSFR $<0.1$\,Gyr$^{-1}$ \citep{mcl18}. Over the redshift range of interest for this study, the VANDELS galaxies are largely sampled from the star-forming main sequence.

The latest VANDELS data release, DR4 \citep{gar21} \footnote{\url{http://vandels.inaf.it/dr4.html}} contains spectra of $\sim2100$ galaxies in the redshift range $1.0<z<7.0$, with on-source integration times ranging from 20 to 80 hours, where $>70\%$ of the targets have at least 40 hours of integration time. The survey depths have been chosen so that the final spectra have high signal-to-noise ratios (SNRs), enabling detailed absorption and emission line studies, the study of ionising sources using emission line ratios, the derivation of accurate metallicities and better constraints on physical parameters such as stellar masses and star-formation rates. The spectral resolution of the final spectra is $R\sim600$. 

The spectra used in this study were reduced using the \textsc{Easylife} data reduction pipeline \citep{gar12} and the final data products delivered by VANDELS include the extracted 1D spectra, the 2D re-sampled and sky-subtracted spectra and catalogues with physical parameters, including spectroscopic redshifts. The reliability of redshifts in the VANDELS database is recorded using the following flags: 0 -- no redshift could be assigned, 1 -- 50\% probability to be correct, 2 -- 70-80\% probability to be correct, 3 -- 95-100\% probability to be correct, 4 -- 100\% probability to be correct and 9 -- spectrum shows a single emission line. The typical accuracy of spectroscopic redshift measurements is $\sim150$ km~s$^{-1}$. Details about the final VANDELS data products, redshift determination processes and redshift reliability flags assignment can be found in \citet{pen18}.

In the redshift range $3.11 < z < 3.53$, we identified 90 galaxies with spectroscopic flags $3$ and $4$, which guarantees a $>95\%$ probability that the spectroscopic redshift is correct and reliable. The VANDELS team has performed aperture matched, PSF-homogenised photometry for every source identified in the CANDELS footprint in GOODS-S across available imaging data ranging from UV to NIR wavelengths \citep{mcl18}, which we refer to as the `master' catalogue. Accurate and homogenised photometry is essential for robust spectral energy distribution (SED) fitting to determine key galaxy properties and we rely on photometry from this master catalogue for this study. 

Therefore, moving forward we require every source selected from other publicly available spectroscopic data that are described below to have a matched entry in the master catalogue and below we list the other publicly available data sets from which sources with reliable spectroscopic redshifts were queried.

\subsubsection{VUDS}
The VIMOS Ultra-Deep Survey (VUDS; \citealt{lef15})\footnote{\url{http://cesam.lam.fr/vuds/}} is a survey of $\sim10000$ faint galaxies in the redshift range $2<z\le6$ with $i$-band magnitudes down to $i \approx 27.0$\,AB. The spectroscopic targets were also selected based on either reliable photometric redshifts or the presence of a Lyman break (LBGs) at higher redshifts, similar to VANDELS.

From the VUDS survey, we select sources in the GOODS-S/CDFS field with redshift reliability greater than 95\%, where the redshift reliability flags are determined in a similar manner to VANDELS. We refer the reader to \citet{lef15} and \citet{tas17} for more information about source selection, data reduction and redshift measurements. In total we select 20 sources with reliable spectroscopic redshifts from VUDS.

\subsubsection{VIMOS10}
The publicly available VLT/VIMOS spectroscopic survey in GOODS-S (\citealt{bal10}; VIMOS10) contains spectra obtained using both the Low Resolution and Medium Resolution grisms for galaxies up to $z\sim3.5$ with $B$ and $R$ magnitudes $\sim24-25$\,AB. The spectroscopic targets were predominantly LBGs, similar to VANDELS and VUDS. We only use sources that have `secure' redshifts, i.e., $>99\%$ confidence of being correct owing to the relatively lower integration times of this survey, and find that 43 sources satisfied our selection criteria.

\subsubsection{MUSE-WIDE Survey}
In addition to photo-z or LBG selected spectroscopic targets, we query reliable spectroscopic redshifts from the publicly available MUSE-WIDE survey in the GOODS-S/CDFS field\footnote{\url{https://musewide.aip.de/project/}}, which is a `blind' integral-field unit (IFU) survey carried out without photometric pre-selection \citep{urr19}. The unique aspect of surveys enabled by MUSE\footnote{\url{https://www.eso.org/sci/facilities/develop/instruments/muse.html}} on the VLT is that no photometric pre-selection is required prior to obtaining a spectroscopic redshift. Therefore, spectroscopically confirmed sources are often blindly selected, owing to their emission lines alone and not necessarily their photometric redshifts or Lyman (or Balmer) breaks, which may bias the target selection. The MUSE-WIDE Survey is capable of obtaining spectroscopic redshifts for $98\%$ of photometrically selected galaxies brighter than $24.0$\,AB in \textit{HST}/F775W. This blind galaxy sample is therefore highly complementary to the photo-$z$ or LBG selection implemented by other data sets used in this study.

From this survey, we select sources with redshifts $3.11 < z < 3.53$ and reliable spectroscopic redshifts, i.e. a `confidence' value of 2 or 3, corresponding to an expected error rate of $\lesssim10\,\%$ in the correct identification of emission and absorption features \citep{urr19}. We find that 44 galaxies in the MUSE-WIDE data release satisfy our selection criteria.

\subsubsection{MUSE-HUDF Survey}
We also use the published spectroscopic redshifts from the MUSE deep (coverage area of $3'\times3'$) and ultra deep (coverage area of $1'\times1'$) surveys in the Hubble Ultra Deep Field (HUDF; \citealt{ina17})\footnote{\url{http://muse-vlt.eu/science/udf/}}, which is part of the GOODS-S field. This survey is deeper than MUSE-WIDE and is capable of blindly detecting sources as faint as $\approx30.0$\,AB in \textit{HST}/F775W \citep{ina17}. 

We find that 42 sources from MUSE-HUDF lie in the redshift range of interest and have reliable spectroscopic redshifts, represented by confidence levels 2 and 3 and these sources are added to our sample.

\subsubsection{3D-HST}
Additional redshifts were queried from the 3D-HST survey, which is a spectroscopic survey conducted using the WFC3/G141 as well as the ACS/G800L grisms on the \emph{HST} covering $\sim600$ square arcminutes of well-studied extragalactic fields, including GOODS-S \citep{bra12, mom16}. The 3D-HST survey obtained blind spectroscopy of galaxies, with the redshift determination primarily relying on the \oii\ emission line in the redshift range of interest to this study. Since there was no photometric pre-selection required, the 3D-HST survey is similar to the MUSE surveys. 

We add a further 9 sources to our study from 3D-HST that have ``good'' redshifts (\texttt{use\_subj = 1}) from the the publicly available catalogues\footnote{\url{https://3dhst.research.yale.edu/Data.php}}.

\subsubsection{Source duplication and final sample}
In total we shortlist 248 sources across the above-mentioned surveys for this study, targeting galaxies based on photometric redshifts of Lyman-break selection, which we refer to as the Lyman break or LBG-selected, and those that were `blindly' detected without any photometric pre-selection, which we refer to as the `blind selected' sample. We note here again that although the spectroscopic surveys have different target selection strategies, we eventually require a detection in the PSF-homogenised CANDELS master photometric catalogue in GOODS-S for every source going forward.

Due to overlapping areas covered by the various spectroscopic surveys, a low fraction sources have been observed by multiple surveys. In such cases we give the highest precedence to the redshift obtained by VANDELS. In cases where the multiply detected source was not part of VANDELS, we give preference to the redshift recorded by the survey that is most sensitive out of the overlapping surveys. 

Removing duplicate observations, 219 unique sources remain in our final sample and we show the breakdown of sources from various spectroscopic surveys in Table \ref{tab:data}. We now turn our attention to filtering out sources that may suffer contamination to their LyC flux by emission from foreground sources along the line-of-sight using high-resolution \textit{HST} photometry, as well as identifying possible AGN in the sample using deep X-ray data.
\begin{table}
    \centering
    \caption{Number of sources with spectroscopic redshifts from data sets used in this study, at every stage of the selection described in \S\ref{sec:sample}.}
    \begin{tabular}{l c c}
    \hline
    Data set & $N$ & Reference \\
    \hline
    \emph{Lyman-break (LBG) selected} \\
    VANDELS DR4 & 90 & (1, 2)\\
    VUDS & 20 & (3, 4)  \\
    VIMOS10 & 43 & (5) \\
    \emph{Blind selected} \\
    MUSE-WIDE & 44 & (6) \\
    MUSE HUDF & 42 & (7) \\
    3D-HST & 9 & (8) \\ \\
    
    Total & 248 & $-$ \\
    \textbf{Total unique} & \textbf{219} & $-$ \\ \\

    Possible contaminants & 34 & $-$ \\
    Possible AGN & 2 & $-$ \\ \\

    \textbf{Final sample} & \textbf{183} & $-$ \\
    Total LBG selected  &  111 & $-$ \\
    Total blind selected & 72 & $-$ \\
    \hline \\
    \end{tabular}
    
    References: (1) \citet{mcl18}; (2) \citet{pen18}; (3) \citet{lef15}; (4) \citet{tas17}; (5) \citet{bal10} (6) \citet{urr19}; (7) \citet{ina17}; (8) \citet{mom16}.
    \label{tab:data}
\end{table}

\subsection{Identifying possible foreground contaminants}
\begin{figure*}
    \centering
    \begin{minipage}{\textwidth}
    \includegraphics[width=\textwidth]{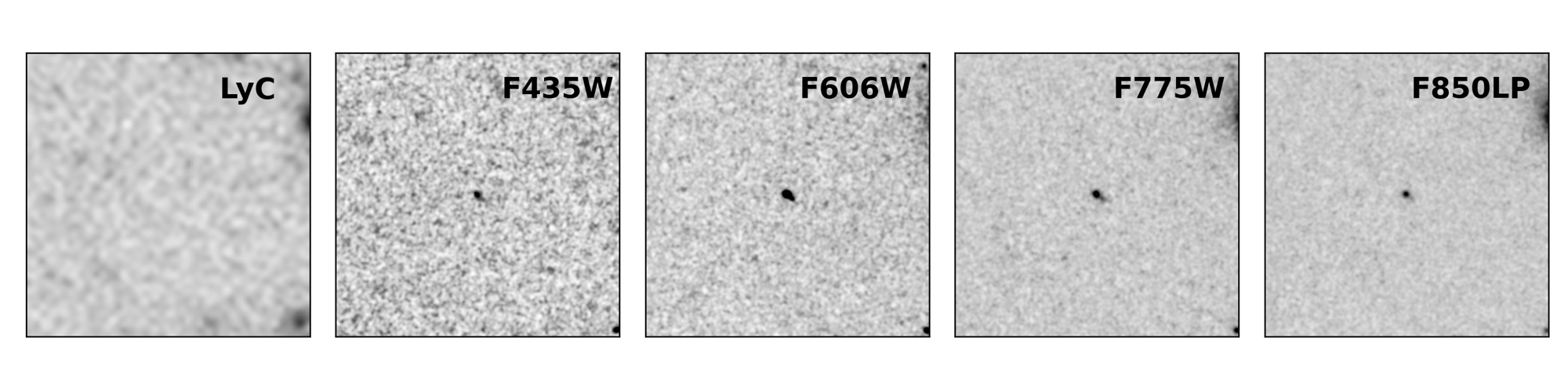}
    \end{minipage}
    \begin{minipage}{\textwidth}
    \vspace{-15pt}
    \includegraphics[width=\textwidth]{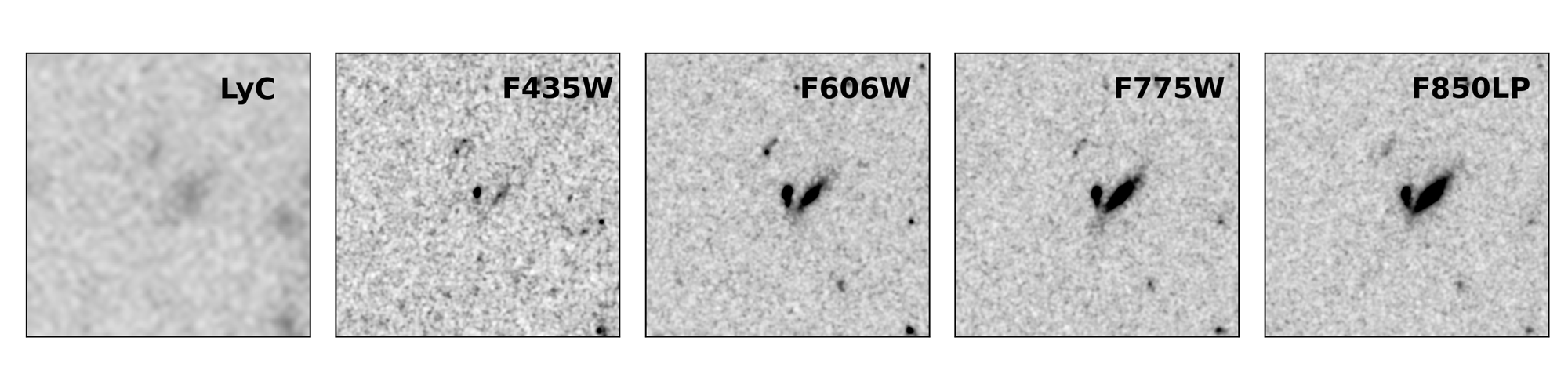}
    \end{minipage}
    \vspace{-15pt}
    \caption{Ground-based LyC and \textit{HST} postage stamps ($5''\times5''$) for a target galaxy that is isolated (top) and a target galaxy that has a possible foreground contaminant within \SI{1}{\arcsec} (bottom). For galaxies that are likely contaminated by foreground objects, examining the colours of both the target galaxy and the contaminant aids considerably in identifying photometric contamination in the LyC bands by a foreground galaxy (see Figure \ref{fig:companion_colour}) that is often mistakenly interpreted as LyC signal \citep[e.g.][]{mos15, gra16, gua16}.}
    \label{fig:example_isolated}
\end{figure*}

\begin{figure}
    \centering
    \includegraphics[scale=0.44]{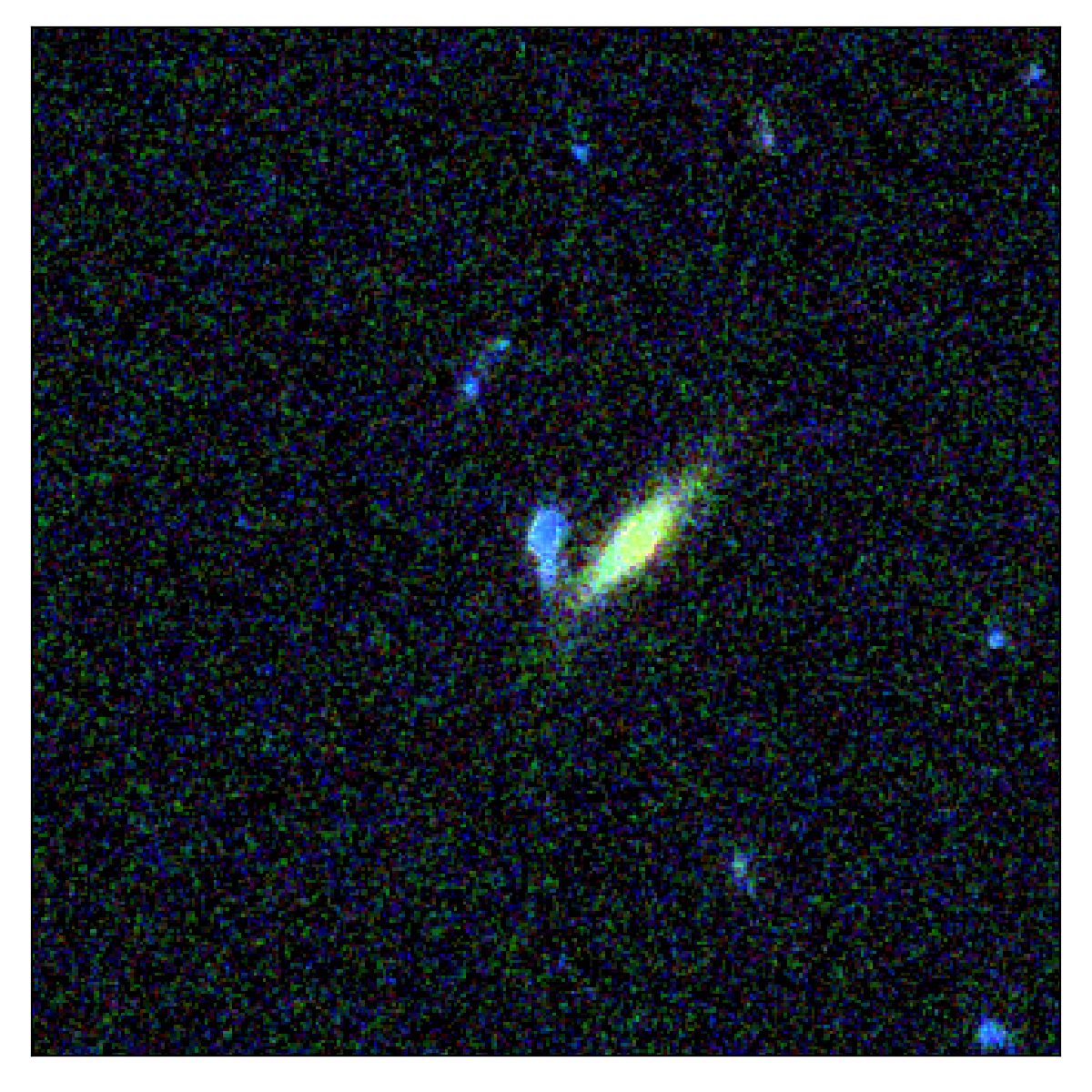}
    \caption{Colour image ($5''\times5''$) using the F606W (blue), F775W (green) and F850LP (red) \emph{HST} filters of the central and companion galaxy of the example shown in the bottom panel of Figure \ref{fig:example_isolated}. The colour of the companion is clearly different from the central galaxy that is of interest, which indicates that the companion source is likely a foreground object.}
    \label{fig:companion_colour}
\end{figure}

One of the biggest sources of error in the measurement of LyC signal from intermediate redshift galaxies is contamination within the aperture from emission from foreground galaxies along the line-of-sight to the source of interest \citep[e.g.][]{van10a}. Therefore, before proceeding to measure LyC fluxes it is important to identify and remove sources that suffer from possible contamination to their LyC signal in the form of image artefacts, bright neighbouring sources or low-redshift interlopers along the line-of-sight. We adopt a series of steps, similar to \citet{gua16}, to filter the sample in order to maximise the accuracy of the LyC signal measurement.

Since all sources in our sample lie within the footprint of the GOODS-S field, they benefit from deep \emph{HST} photometry at optical and near-infrared wavelengths from the \emph{HST} Treasury Program \citep{dic03} and through CANDELS \citep{gro11, koe11}, offering unparalleled resolution. This resolution is critical in the identification of possible foreground neighbouring sources in the vicinity of the galaxies with spectroscopic redshifts \citep[e.g.][]{mos15} that may strongly contaminate any LyC signal observed from the relatively lower resolution ground-based images. 

To identify potential contamination, we create \SI{20}{\arcsec} $\times$ \SI{20}{\arcsec} postage stamps from images in the F435W, F606W, F775W and F850LP \emph{HST} filters using publicly available images\footnote{\url{https://archive.stsci.edu/pub/hlsp/goods/v2/h_goods_v2.0_rdm.html}}. We also create postage stamps of the same dimensions in the LyC images. Sources that appear to be isolated in the \emph{HST} images, without any other component or a nearby source within \SI{1}{\arcsec} of their centroid coordinates are shortlisted for LyC signal measurement without any further filtration. An example of such a source (with postage stamp sizes of $5''\times5''$ for clarity) is shown in the top panel of Figure \ref{fig:example_isolated}.

For sources that have one or more components within a radius of \SI{1}{\arcsec} of the centroid in the \emph{HST} images, we must determine whether these are a part of the galaxy in question (which may have a `clumpy' morphology) or whether they are possibly foreground objects \citep[e.g.][]{gra16}. In the bottom panel of Figure \ref{fig:example_isolated} we show an example (with $5''\times5''$ postage stamps for clarity) where the spectroscopically confirmed source is surrounded by a nearby source that lies within \SI{1}{\arcsec}. Following \citet{gua16}, we examine the colours of the central galaxy and the surrounding components to identify contaminants in the following way. Using the F435W ($B$), F606W ($V$), F775W ($i$) and F850LP ($z$) images, we generate BGR colour images using the \textsl{make\_lupton\_rgb} function of \textsl{astropy.visualization} in $BVi$, $BVz$, $BIz$ and $VIz$. Additionally, we also produce a $V-z$ image for all sources. By then comparing the colours of the spectroscopically confirmed target and its companion(s) within \SI{1}{\arcsec}, we determine whether the component(s) close to the centroid of the source is (are) part of the system or not. For example, Figure \ref{fig:companion_colour} shows the $Viz$ colour image for the source shown in the bottom panel of Figure \ref{fig:example_isolated}, which is indicative that the companion source is likely to be a foreground contaminant.

\citet{gua16} showed that the above-mentioned simpler and more computationally efficient colour-based foreground identification method is highly effective at identifying interlopers. Such an analysis also returns a comparable contamination identification fraction compared to a full SED-fitting based analysis of the individual components surrounding the spectroscopically confirmed source. We also stress that removing sources that show clumps of different colours is a highly conservative approach, as it might result in the removal of galaxies with clumpy morphologies where the individual clumps may harbour slightly different stellar populations.

Through this process, we remove 34 sources from our initial sample of 220 unique sources, representing $\approx15\%$ of the sample and bringing the sample size of unique sources that are unlikely to suffer from foreground contamination to 186. The fraction of sources removed due to possible contamination by foreground objects is highly consistent with an expected fraction of $12.6\%$ for $z\gtrsim3$ galaxies determined by \citet{van10a}, down to the photometric depths considered in this study assuming a \SI{1}{\arcsec} radius for contaminant identification. 

\subsection{Removing possible AGN}
Finally, we remove potential AGN from our sample by cross-matching the coordinates of our sources with the \emph{Chandra} 7 Ms source catalogue compiled by \citet{luo17} using a matching radius of \SI{1.0}{\arcsec}. The sources detected in the \citet{luo17} catalogue at $z\gtrsim3$ are very likely to be AGN, owing to X-ray luminosities of $>10^{42}$\,erg\,s$^{-1}$ at these redshifts \citep[e.g.][]{mag20, sax20b, sax21}. We identify and remove two sources with counterparts in the X-ray catalogue that are also classified as AGN in the catalogue and therefore, the total number of sources in our final sample comes to 183.

\subsection{Final sample}
After removing duplicates, sources with foreground contamination and AGN from our sample, the total number of unique galaxies in the redshift range $3.11 < z < 3.53$ with reliable spectroscopic redshifts is 183.

The median spectroscopic redshift of our sample is $z=3.38$ and our sample includes the known over-density of star-forming galaxies at $z\sim3.5$ in GOODS-S reported by \citet{for17}. The median \textit{HST}/F606W band magnitude of the galaxies in the final sample is $25.69 \pm 0.84$. We now proceed to fit spectral energy distribution (SED) models to the photometric data available to derive key galaxy physical parameters.

\section{SED fitting and physical properties}
\label{sec:sed}
Since the main aim of this study is to explore the dependence of LyC \fesc\ on key galaxy properties such as stellar masses, star-formation rates, etc. in addition to statistically exploring the correlations between \fesc\ and the strength of rest-frame nebular emission lines tracing intense star-formation such as \ohb, obtaining robust and reliable spectral energy distribution (SED) fits using high quality photometry of galaxies in our sample is crucial. 

These key galaxy properties are potentially important factors that govern \fesc\ from reionisation era galaxies, and in this section we use deep \emph{HST} and \emph{Spitzer}/IRAC photometry in the observed frame optical and near infrared bands, complemented by deep ground based $K$ band photometry in combination with reliable spectroscopic redshifts to accurately constrain the stellar population properties and dust extinction for the 183 sources in our final sample. We additionally compare the derived physical properties of galaxies in our sample that were LBG selected, and those that were blind selected, exploring any inherent differences in galaxies that these two selection techniques may introduce, thereby impacting the inferred \fesc.

\subsection{SED derived physical properties}
\label{sec:sed-fitting}
To determine both the properties of stellar populations as well as the \ohb\ line strengths using $K$ band photometry, we simultaneously fit the stellar continuum and nebular line emission to the observed broadband photometry of galaxies in our sample. We note that since both stellar continuum and nebular emission are being simultaneously fitted, we include the line-contaminated $K$ band photometry in the fitting.
\begin{figure*}
    \centering
    \includegraphics[width=0.75\textwidth]{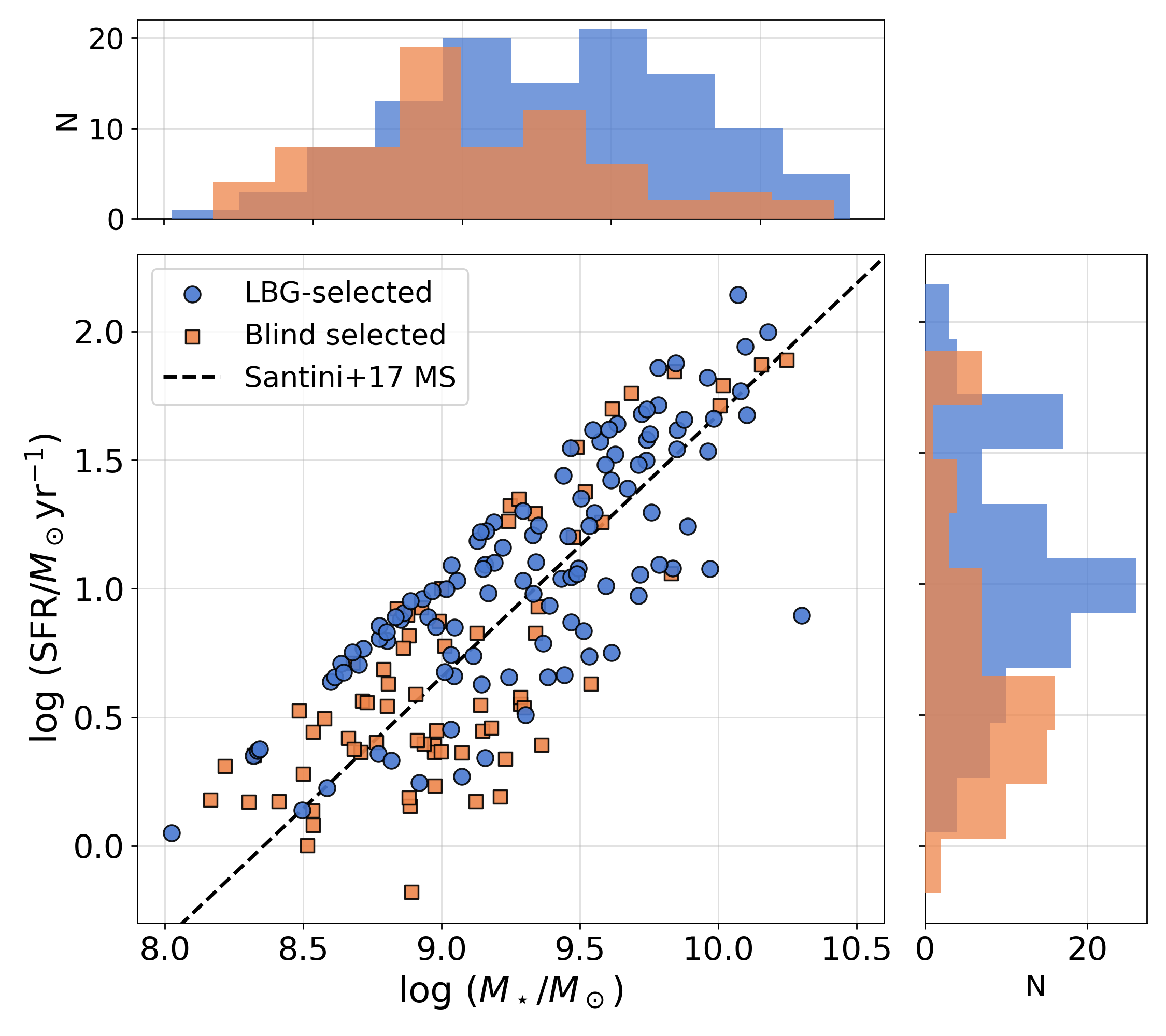}
    \caption{Distribution of SED derived stellar masses and star-formation rates for galaxies in the final sample, colour coded by metallicity. The dashed line shows the star-forming main sequence derived by \citet{san17} at redshifts $3<z<4$. We find that most of our galaxies lie on the main sequence, and span a range of masses and star-formation rates, which makes them representative of the general galaxy population at $z\sim3.3$}
    \label{fig:mass_sfr}
\end{figure*}

The SED fitting is performed using the Bayesian Analysis of Galaxies for Physical Inference and Parameter EStimation (\textsc{Bagpipes}) code \citep{car18}. \textsc{Bagpipes} is a \textsc{python} based tool to generate complex galaxy model spectra, allowing flexible fitting to photometric data to derive key galaxy physical properties. As mentioned earlier, we use a combination of stellar continuum and nebular emission when fitting the SEDs. The stellar models are based on \citet{bc03}, constructed using the \citet{kro02} initial mass function (IMF). The nebular emission is computed using Cloudy grids \citep{fer13}, with the only free parameter being the log of the ionisation parameter ($\log U$). When determining the strengths of the emission lines from these grids, the metallicity of the line-emitting gas is assumed to be the same as that of the stars.

We assume an exponentially declining star-formation history ($\tau$-model), with stellar ages allowed to vary between 10\,Myr and the age of the Universe at the observed redshift of the galaxy. The value of $\tau$ is allowed to vary between 100\,Myr to 10\,Gyr, with stellar metallicities allowed to vary in the range $0-1\,Z_\odot$, where $Z_\odot$ is the solar metallicity. The \citet{cal00} dust attenuation law is applied and the dust extinction, $A_{1400}$ is allowed to vary between 0 and 2 mag. We use flat posterior distributions for each of the parameters, which is the default choice in \textsc{Bagpipes}. The nebular emission line strengths are set by the $\log U$ parameter, which we allow to vary from $-3$ to $-2$, which are the typically inferred values for $z\gtrsim3$ star-forming galaxies \citep[see][for example]{nak18, sax20}. 

The chi-squared values of the best-fitting SEDs are evaluated, and individual SEDs are then visually inspected to ensure that the broadband photometry, and crucially the $K$ band flux, are well reproduced by the best-fitting model. The best-fitting SEDs are then used to infer key physical properties for each galaxy, such as stellar age, dust attenuation, stellar mass, metallicity, star-formation rates, etc., in addition to the fluxes and rest-frame equivalent widths of the \oiii\,$\lambda5007$ and the \hb\ lines derived from the nebular component of the best-fitting SED. Thanks to the Bayesian inference methodology of \textsc{Bagpipes}, posterior distributions for all physical parameters are derived, leading to accurate uncertainty estimates.

The strength of the nebular emission that is added to the stellar continuum using \textsc{Bagpipes} crucially depends on the best-fitting metallicity as well as the ionisation parameter \citep[see][]{car18}. Although the ionisation parameter is a variable that can be controlled, the best-fitting metallicities are inferred from the broadband SED fitting using both stellar and nebular components.

The median stellar mass (with $1\sigma$ standard deviation) derived from SED fitting of all 183 galaxies is $\log(M_\star/M_\odot) = 9.2 \pm 0.5$, the median star-formation rate is $8 \pm 24$\, M$_\odot\,\textrm{yr}^{-1}$ , the median stellar age is $1.5 \pm 1.4$\,Gyr, and the median stellar metallicity is $0.6 \pm 0.3$\,$Z_\odot$, where $Z_\odot$ is the solar metallicity \citep{asp09}. Overall, the population of galaxies considered in this study are consistent with low-mass, star-formation objects with young stellar ages and low stellar metallicities. 

In Figure \ref{fig:mass_sfr} we show the distribution of the derived stellar masses and star-formation rates of all the galaxies in the final sample. We mark the LBG-selected galaxies in our sample using blue points, and the emission line selected galaxies using orange squares. The histograms of both stellar masses and SFRs for the LBG and blind selected samples are also shown. We show the star-forming main sequence at $3<z<4$ determined by \citet{san17} using dashed lines.

The majority of galaxies in our sample lie on or around the main sequence, and that the stellar masses and star-formation rates vary over a large range. We also find that as expected, the blind selected galaxies tend to occupy the lower stellar mass end of the main-sequence, consequently also exhibiting lower star-formation rates compared to LBG-selected galaxies.

We note here, however, that owing to the magnitude limited nature of our galaxy sample, a larger fraction of galaxies with lower stellar masses detected in the spectroscopic surveys will cluster towards higher SFRs, as is evident from Figure \ref{fig:mass_sfr}, thereby resulting in an overall higher sSFR probed at lower stellar masses. Such an effect can potentially bias any possible dependence of LyC \fesc\ on key galaxy properties such as stellar mass and/or sSFR.

\subsection{\ohb\ line strengths}
\label{sec:eelg-id}
\begin{figure}
    \centering
    \includegraphics[scale=0.4]{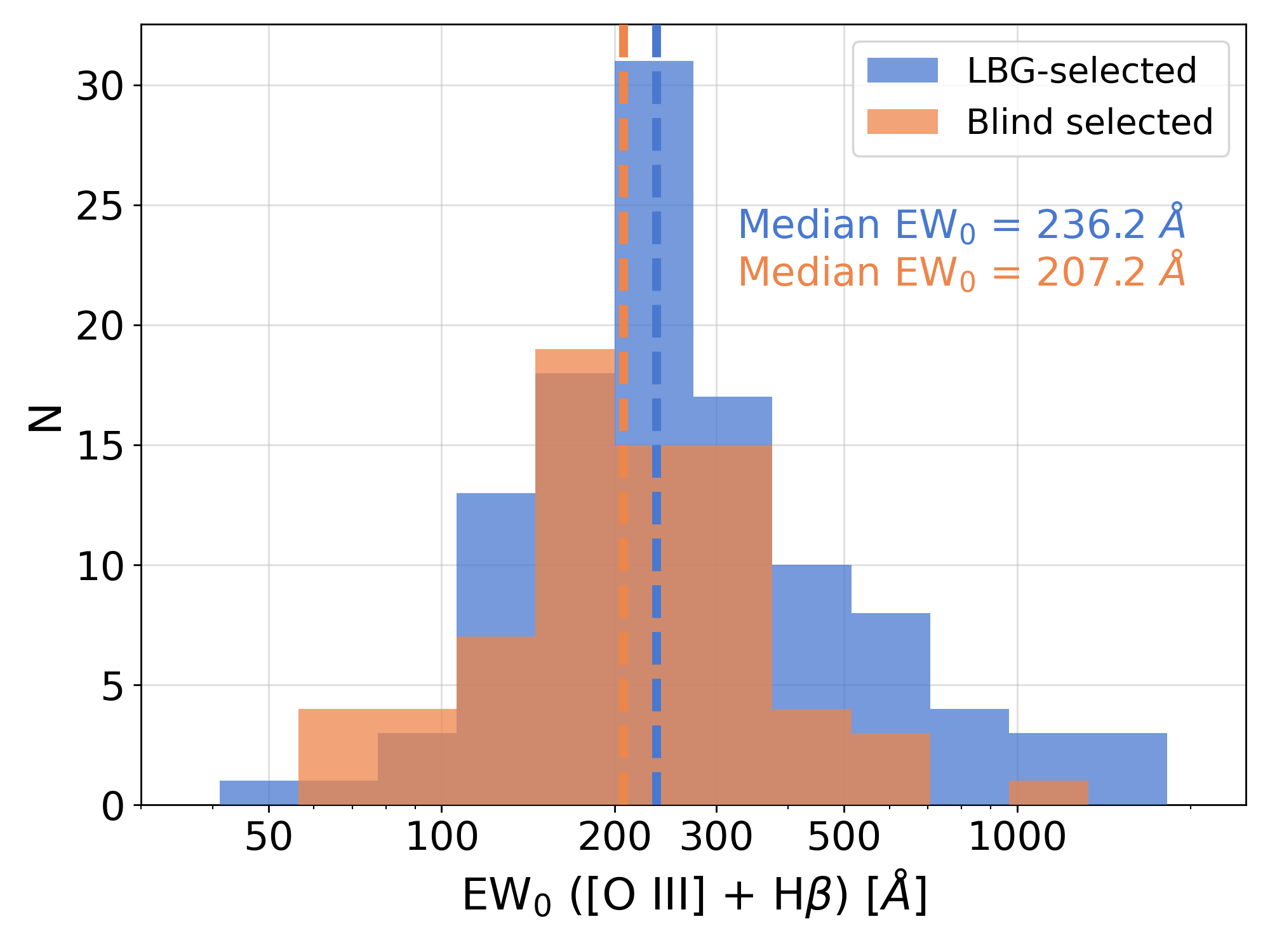}
    \caption{Histogram of the rest-frame \ohb\ equivalent widths measured for LBG-selected galaxies (blue) and blind selected galaxies (orange) in our sample using SED fitting. The median EW$_0$(\ohb) of the full sample is $228.6$\,\AA, and the derived equivalent widths follow a log-normal distribution (similar to that reported by \citealt{sta13}. The EW distributions for both LBGs and emission line galaxies are comparable, with the median values (dashed lines and text) highly consistent with each other.}
    \label{fig:oiii_strengths}
\end{figure}

The best-fitting SED includes a stellar continuum and a nebular emission line component, from which the equivalent width of the emission lines can be calculated. Particularly for the \oiii\ and \hb\ lines that fall in the observed $K$ broad band, we measure a median EW$_0$(\ohb) of $\approx 230$\,\AA\ with standard deviation $225$\,\AA, with the LBG selected sample having a median EW$_0$(\ohb) $= 236.2$\,\AA\ and the blind selected sample having a slightly lower median EW$_0$(\ohb) $= 207.2$\,\AA. The distribution of EW$_0$(\ohb) measured for LBGs and blind selected galaxies is shown in Figure \ref{fig:oiii_strengths}. 

\begin{figure}
    \centering
    \includegraphics[scale=0.4]{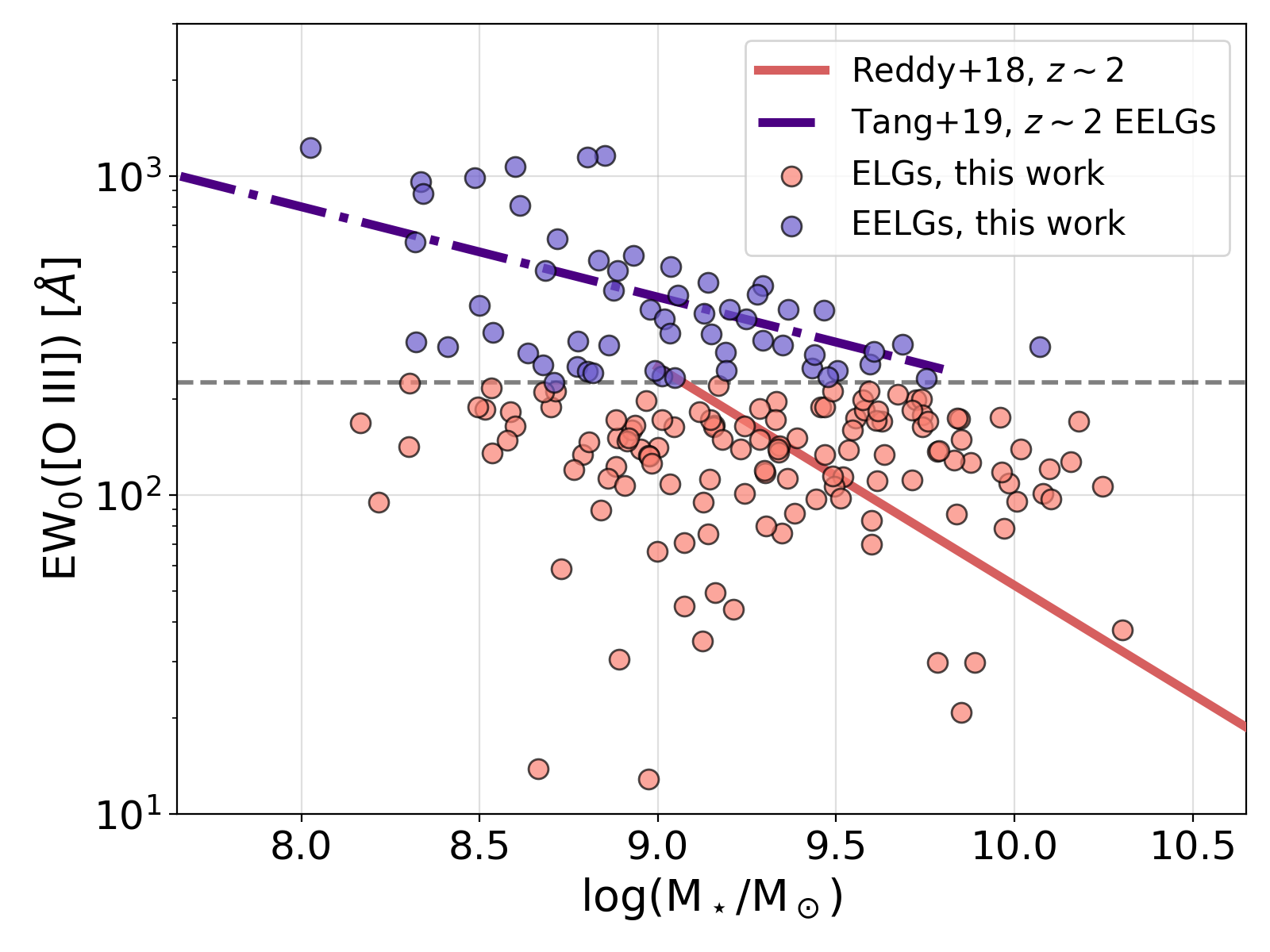}
    \caption{Distribution of EW$_0$(\oiii) and galaxy stellar mass derived for our sample of galaxies using SED fitting. Shown for comparison are the best-fitting relations derived between the two quantities for relatively massive star-forming galaxies at $z\sim2$ by \citet{red18} (red solid line) from the MOSDEF survey and extreme emission line galaxies (EELGs) at $z\sim2$ by \citet{tan19} (blue dot-dashed line), where the black dashed line marks EW$_0$(\oiii) $>225$\AA, used by \citet{tan19} to classify galaxies as EELGs. The normalisation of the best-fitting relations have not been altered. Galaxies from our sample that would be considered as normal emission line galaxies (ELGs) are shown as red points, and those considered as EELGs are shown using blue points. The distribution of our SED derived measurements (and the scatter) are broadly consistent with spectroscopic measurements at lower redshifts.}
    \label{fig:oiii-ew-comparison}
\end{figure}

The most immediate comparison for our \ohb\ equivalent widths can be made with the study of \citet{for17}, where using $K$ band photometry for galaxies at $2.95 < z < 3.65$ they report an average EW$_0$(\ohb) $=230 \pm 90$\,\AA, which is highly comparable to the median value we find for our sample. In Figure \ref{fig:oiii-ew-comparison} we compare our SED derived line measurements with the best-fitting relations derived between spectroscopic measurements of the \oiii\ $\lambda5007$ line and the stellar masses of a large sample of star-forming galaxies at $z\sim2$ as part of the MOSDEF survey \citep{red18}. We additionally show the stellar mass dependence of the \oiii\ line derived for a sample of extreme emission line galaxies (defined as EW$_0$(\oiii) $>225$\AA\ by \citealt{tan19}) at $z\sim2$ by \citet{tan19}, which also exhibit lower stellar masses.

We find that the distribution of EW$_0$(\oiii) as a function of stellar mass for galaxies in our sample broadly agrees with the best-fitting relation derived from MOSDEF at higher stellar masses. The scatter around the best-fitting relation observed in our sample is also consistent with what was reported by \citet{red18}. 

We identify 59 galaxies in our sample that can be classified as extreme emission line galaxies (EELGs), with EW$_0$(\ohb)~$>300$\,\AA\ \citep[e.g.][]{star16}). The median EW$_0$(\ohb) in the EELG sample is $\approx500$\,\AA, which is slightly lower than the range of $\approx 660-800$\,\AA\ reported by \citet{for17} for EELGs in their sample. 

To further investigate the reliability of our inferred \ohb\ line strengths, we compare our rest-frame \ohb\ equivalent width measurements as a function of galaxy stellar to results from the MOSDEF spectroscopic survey \citep{kri15}, which spectroscopically observed a sample of $\sim1200$ star-forming galaxies at redshifts $1.4\lesssim z \lesssim 3.8$ \citep{red18} in Figure \ref{fig:oiii-ew-comparison}. We find that our measurements at $z\sim3-3.5$ broadly agree with the best-fitting relation obtained by \citet{red18}, with comparable scatter at stellar masses $>10^9$\,$M_\odot$.

We also compare EWs of galaxies in our sample that may be considered as EELGs following the \citet{tan19} classification (EW$_0$(\oiii) $>225$\AA) with spectroscopic measurements at $z\sim2$ from \citet{tan19}, finding broad agreement with their best-fitting relation (Tang et al., private communication), which is also shown in Figure \ref{fig:oiii-ew-comparison}.

Our EELG \ohb\ EWs are also comparable with photometrically selected extreme line emitters found at $z>6$ -- \citet{end20} measured a median value of EW$_0$(\ohb) $\sim700$\,\AA\ for star-forming galaxies at $z\sim7$ and \citet{lab13} found an average rest frame \ohb\ EW of $670\pm200$\,\AA\ at $z\sim8$. Additionally, \citet{smi15} identified galaxies with extreme Spitzer/IRAC colours at $z\sim6.8$ and found that 50\% of the strongest line emitters have EW$_0$(\ohb) $\sim1085$\,\AA. Using a similar technique, \citet{rob16} identified four bright galaxies at $z\sim7$ with EW$_0$(\ohb)\ $\sim1500$\,\AA, all of which have been confirmed to be strong \lya\ emitters. Further, \citet{cas17} measured an average EW(\ohb) of $\sim1500$\,\AA for \lya\ emitting galaxies at $z\sim6.8$. Therefore, our sample of EELGs at $z\sim3-3.5$ offers an excellent opportunity to study the escaping ionising radiation from analogues of reionisation-era galaxies.

We note here, however, that the line strengths of some of the brightest \ohb\ emitting galaxies, particularly those with fainter continua, may be underestimated due to the lower limit on the ionisation parameter of $-2.0$ permitted by \textsc{Bagpipes} as well as the depth of photometric data in the $K$ band. We further note that the flux (and consequently stellar mass) limited nature of our galaxy sample naturally results in incompleteness of EELGs down to lower stellar masses.


\subsection{Comparison with NIR spectroscopic observations}
\label{sec:nirvandels-sed}
To test the accuracy of the inferred \ohb\ strengths from SED fitting, we compare our measurements with near-IR observations of a subset of VANDELS galaxies in the CDFS and UDS fields in the redshift range $z\sim3-4$ using MOSFIRE on Keck Telescope, as part of the recently completed NIRVANDELS survey \citep{cul21}. We note that only a fraction of NIRVANDELS galaxies in the CDFS field are also part of the final sample in this study.

\begin{figure}
    \centering
    \includegraphics[scale=0.4]{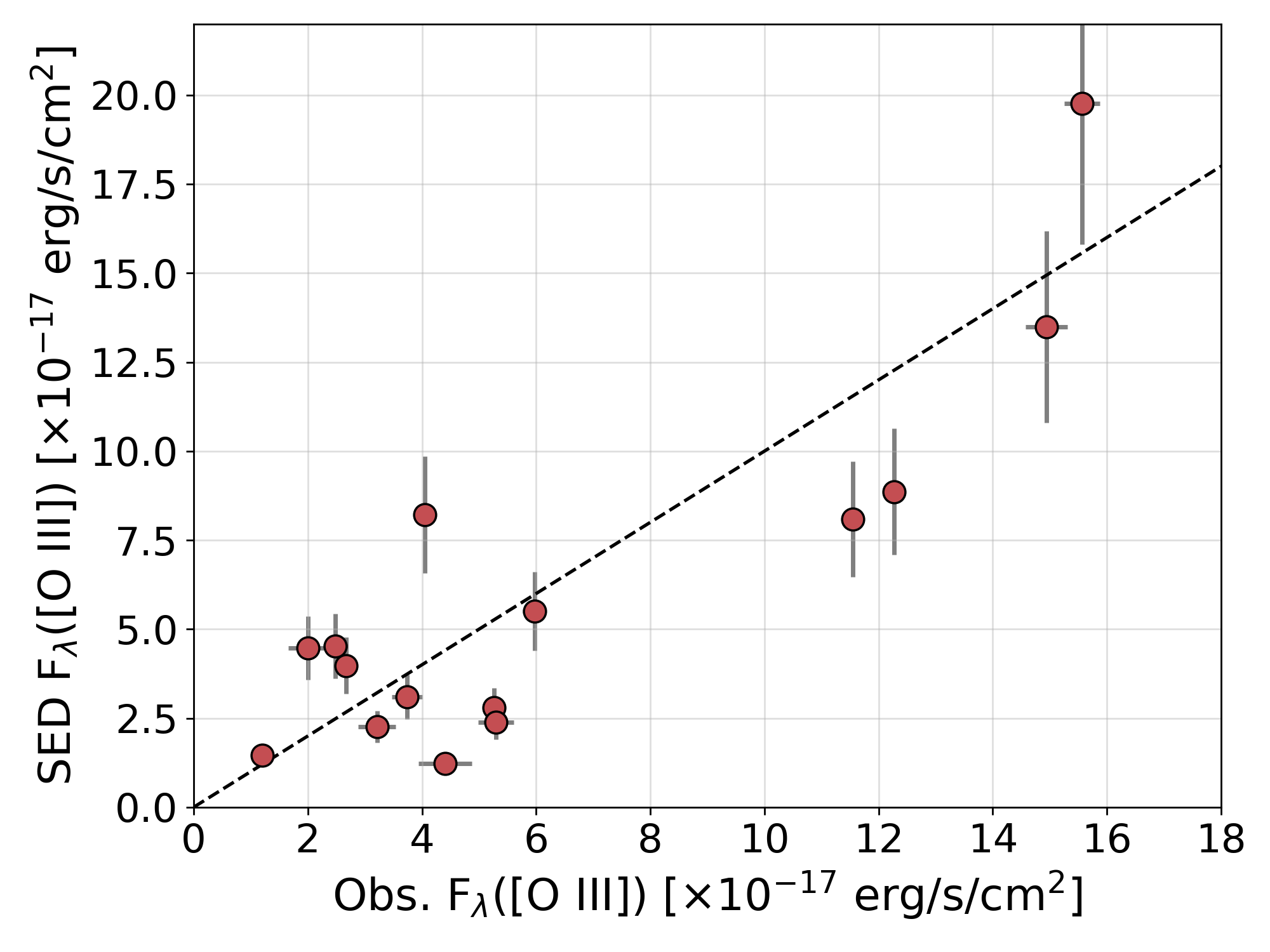}
    \caption{Comparison of \ohb\ line fluxes measured from the best-fitting SED and $K$-band excess and the observed line fluxes from NIR spectra. From the limited number of galaxies in our sample that have observed line fluxes from \citet{cul21}, we can conclude that our method of estimating line fluxes from $K$-band excess works reasonably well.}
    \label{fig:kband-mosfire}
\end{figure}

To do this, we fit SEDs to all NIRVANDELS galaxies using the same SED fitting parameters and techniques that we used for the main galaxy sample described in \S\ref{sec:sed} and compare the observed and SED-derived \oiii\ line fluxes. The comparison between the SED derived and observed \oiii\ line fluxes for NIRVANDELS galaxies is shown in Figure \ref{fig:kband-mosfire}, with the dashed line marking the one-to-one relation. The error bars shown on the SED derived line fluxes are at the 20\% level. We find that the SED derived line fluxes are remarkably close to the observed values (within $\sim0.3$\,dex) over a broad flux range. We therefore conclude that simultaneously fitting the stellar continuum and nebular line emission with the ionisation parameter in the range $-3 \leq \log U \leq -2$ gives realistic estimates of the \oiii\ (and \hb) line fluxes for galaxies in our sample.

Having fitted SEDs to all 183 unique galaxies in our final sample and measured key galaxy properties as well as \oiii\ and \hb\ line strengths, we now proceed to measure LyC fluxes for galaxies using the available LyC images.

\section{Lyman continuum escape fraction}
\label{sec:lyc}
\subsection{Measuring \fescrel\ and \fesc}
In this section we describe our methodology for measuring the LyC flux from the available LyC images. Since these filters cover slightly different rest-frame LyC wavelengths at different redshifts as mentioned in \S\ref{sec:filters}, we split our sample into the relevant redshift bins corresponding to LyC coverage offered by the filters. \textit{NB3727} includes 79 galaxies, \textit{NB396} includes 83 galaxies, and \textit{U} includes 91 galaxies. We show the number of galaxies and the redshift ranges corresponding to each LyC filters along with $1\sigma$ depths in Table \ref{tab:sources}. We note that owing to the overlapping wavelength coverage offered by the the \textit{NB396} and the \textit{U}-band filter, some galaxies are included in both subsets.
\begin{table*}
    \centering
    \caption{LyC filters used in this study along with $1\sigma$ depths (see \S\ref{sec:filters}), LyC redshift range covered and the number of sources. Note that some sources have overlapping LyC coverage in \textit{NB396} and \textit{U} images.}
    \begin{tabular}{l c c c c c}
    \hline 
    Filter  & Tel./Instr. & $1\sigma$ depth (AB) & $z$ range & N & Reference  \\
    \hline
    \emph{NB3727} & CTIO 4m/MOSAICII & 27.38 & $3.11 - 3.34$ & 79 & \citet{gua16} \\
    \emph{NB396} & La Silla 2.2m/WFI & 27.72 & $3.41 - 3.53$ & 83 & \citet{gua16} \\
    $U$ band & VLT/VIMOS & 29.78 & $3.34 - 3.49$ & 91 & \citet{non09} \\    
    \hline
    \end{tabular}
    \label{tab:sources}
\end{table*}

An important parameter quantifying the escape of LyC photons from a galaxy in a dust-independent manner is the relative \fesc, or \fescrel\ \citep{ste01}, comparing the observed ratio of ionising to non-ionising UV flux to the intrinsically produced ratio by stellar populations within the galaxy, corrected for absorption by the intervening neutral IGM. If the intrinsic ratio of luminosities at $\sim1400$\,\AA\ and $\sim900$\,\AA, $(L_{1400}/L_{900})_{\textrm{int}}$ is known (e.g. from stellar population modelling), then \fescrel\ can be calculated as 
\begin{equation}
\label{eq:fescrel}
    f_{\textrm{esc}}^{\textrm{rel}} = \left(\frac{F_{900}}{F_{1400}}\right)_{\textrm{obs}} \times \left(\frac{L_{1400}}{L_{900}}\right)_{\textrm{int}} \times
    \frac{1}{\exp \left(-\tau_{900}\right)}
\end{equation}
where ${(F_{900}/F_{1400})_{\textrm{obs}}}$ is the observed ratio of the ionising LyC flux to non-ionising UV flux at $\sim1400$\,\AA, and $\tau^\textrm{IGM}_{900}$ is the transmissivity of ionising photons (at $900$\,\AA) through the intervening IGM \citep{sia07}. The absolute escape fraction, \fesc, can then be calculated from \fescrel\ by applying a dust correction as follows
\begin{equation}
\label{eq:fesc}
    f_\textrm{esc} = f_\textrm{esc}^\textrm{rel} \times 10^{(-0.4 \times A_{1400})}
\end{equation}
where $A_{1400} = 10.33 \times E(B-V)$ according to the \citet{cal00} dust reddening law. Therefore, to calculate \fesc\ we must constrain each of these ingredients in equations (1) and (2), and below we highlight how these are calculated for our galaxies.

\subsubsection{LyC flux measurement}
\label{sec:photometry}
The observed flux ratios in the equations above require a measurement of the LyC flux, which we measure from the available LyC images using aperture photometry. Photometry in the LyC images is performed by placing circular apertures at the coordinates in the CANDELS catalogues using the \textsc{python} package \textsc{photutils} \citep{photutils}. The average PSF of the \emph{NB3727} image is \SI{1.4}{\arcsec}, of the \emph{NB396} is \SI{0.94}{\arcsec} \citep{gua16} and of the \emph{U} image is \SI{0.8}{\arcsec} \citep{non09}. The radius of the circular aperture is set to either match or be slightly larger than the PSF to capture small-scale PSF variations across the LyC images. Therefore, we use an aperture with radius \SI{0.7}{\arcsec} for \textit{NB3727} and \SI{0.5}{\arcsec} for \textit{NB396} and \textit{U} band images. 

We measure the local background noise around the apertures by placing a circular annulus (at the same positions as the aperture) with an inner radius of \SI{2}{\arcsec} and an outer radius of \SI{3}{\arcsec}. For galaxies in which the flux measured in the aperture was lower than the background flux measured in the annulus, we assigned a limiting LyC flux value equal to the local background noise with a 100\% error. 

The non-ionising UV flux is calculated from the F606W \textit{HST} image, which covers the rest-frame $\sim1400$\,\AA\ wavelengths for the redshift range of galaxies in this study. Therefore, the ratio of the flux determined from the LyC bands and from the F606W band gives us a measure of the term $(F_{900}/F_{1400})_{\textrm{obs}}$. 

\subsubsection{Intrinsic non-ionising to ionising luminosity ratio}
The intrinsic ratio between UV luminosity at non-ionising wavelengths (e.g. $\sim1400$\,\AA) and LyC wavelengths ($\sim900$\,\AA) is calculated individually for each source using the best-fitting SED. We create synthetic SEDs using the best-fitting stellar population parameters determined and measure the intrinsic ratios, $L_{1400}/L_{900}$, for each galaxy.

We find an average intrinsic ratio of $4.5$, with a standard deviation of $0.7$. Individual intrinsic ratio measurements for each galaxy enabled by accurate spectroscopic redshifts lead to a more accurate \fesc\ calculation for individual galaxies compared assuming an average intrinsic ratio for all galaxies in the sample, as done by previous studies measuring LyC \fesc\ at $z\sim3$, where the assumed intrinsic ratio took values of 3 \citep{ste01, gra16, gra17, mar16}, 5 \citep{gua16, van16a} or 7 \citep{fle19}.

\subsubsection{IGM transmissivity}
We calculate the IGM transmissivity of LyC photons for each galaxy using the prescriptions of \citet{ste18}. In addition to the IGM transmission, the \citet{ste18} model also takes into account the effects of transmission through the circumgalactic medium (CGM) surrounding each galaxy in the so-called ``IGM+CGM'' model. This results in an average LyC transmission, $\exp(-\tau_{900}) = 0.371$ at $z=3.0$ and a much lower $\exp(-\tau_{900}) = 0.264$ at $z=3.5$.

An advantage of having spectroscopic redshifts is that the IGM transmission can be accurately calculated for each source at the rest-frame wavelength of $900$\,\AA, the $\tau_{900}$. Hence, we calculate the LyC transmission through both the IGM and the CGM, or $\exp(-\tau_{900})$ for each galaxy in our final sample.

\subsubsection{Dust attenuation}
Finally, the dust attenuation (or $A_{1400}$) for each source is calculated from the best-fitting SED. During the SED fitting process we allowed $A_{1400}$ to vary between 0 and 2, which naturally limits the derived values. This means that our SED fits will miss any significantly dusty galaxies, but given the high-redshift star-forming nature of these galaxies, heavy dust obscuration is unlikely to be the case. We find an average dust attenuation of $A_{1400} = 0.40$ mag in our sample with a standard deviation of $0.26$ mag.

Having defined the necessary ingredients to calculate \fesc\ for each source and described the measurements of each of these parameters, in the following section we present the our measurements of the relative and the absolute escape fractions of LyC photons from each galaxy.

\subsection{Individual LyC \fesc\ measurements}
\label{sec:results-ind}
We measure non-zero LyC \fesc\ at $\gtrsim2\sigma$ significance levels in 11 out of 183 galaxies in our sample. The LyC \fesc\ is in the range of $0.14-0.85$, and the galaxies showing non-zero \fesc\ span a wide range of stellar masses ($M_\star = 10^{8.3} - 10^{9.7}\,M_\odot$), star-formation rates (SFR $=2.3 - 37.4\,M_\odot$\,yr$^{-1}$) and \ohb\ equivalent widths (EW$_0$(\ohb) $= 111.8 - 1160.3$\,\AA). 

We note that 8 out of 11 candidate LyC leakers were LBG selected, and 3 were blindly selected for spectroscopic follow-up. The galaxy with the highest measured \fesc\ $=0.85\pm0.33$ (CDFS 13385) was blindly selected from the MUSE-HUDF survey and has a relatively low stellar mass of $10^{8.3}\,M_\odot$, showing extreme \ohb\ line emission through its $K$ band photometry.

The best-fitting SED derived physical parameters for these 11 candidate LyC leaking galaxies are given in Table \ref{tab:lyc-physical}, and the measured quantities (both through SED fitting and aperture photometry) used to calculate \fescrel\ and \fesc\ are given in Table \ref{tab:lyc-fesc}.  The $5''\times 5''$ cutout galleries showing the LyC image and the observed-frame optical \textit{HST} photometry, along with the best-fitting SEDs for the 11 LyC leaker candidates are shown in Appendix \ref{appendix:a}. A full spectroscopic analysis of these 11 LyC leaking candidates is beyond the scope of the present paper and will be explored in greater detail in forthcoming publications.
\begin{table*}
    \centering
    \caption{Nature of selection for each galaxy, with LBG representing Lyman-break selected and BLD representing blind selection, as well as best-fitting SED derived physical properties for candidate LyC leaking galaxies. Also given is the LyC detection filter for each source.}
    \begin{tabular}{l c c c c c c c c}
    \hline
    CANDELS ID & Sample & Survey & $z$ & LyC filter & log($M_\star$/$M_\odot$) & SFR (\sfr) & log(sSFR/yr$^{-1}$) & EW$_0$ (\ohb) (\AA) \\
    \hline
    CDFS 5161 & LBG & VANDELS & 3.420 & NB396 & 9.3 & 16.3 & $-8.1$ & 268.2 \\ 
    CDFS 9358 & LBG & VIMOS10 & 3.229 & NB3727 & 9.6 & 37.0 & $-8.1$ & 241.3 \\ 
    CDFS 9692 & LBG & VANDELS & 3.470 & NB396 & 8.3 & 2.3 & $-8.0$ & 1160.3 \\ 
    CDFS 12448 & LBG & VANDELS & 3.460 & U & 9.6 & 12.5 & $-8.5$ & 119.3 \\ 
    CDFS 13385 & BLD & MUSE HUDF & 3.431 & U & 8.6 & 3.6 & $-8.1$ & 334.6 \\ 
    CDFS 15718 & BLD & MUSE HUDF & 3.439 & NB396 & 8.3 & 2.0 & $-8.0$ & 986.8 \\ 
    CDFS 16444 & BLD & MUSE HUDF & 3.128 & NB3727 & 9.3 & 23.1 & $-7.9$ & 456.4 \\ 
    CDFS 18454 & LBG & VUDS & 3.163 & NB3727 & 9.2 & 16.7 & $-8.0$ & 339.8 \\ 
    CDFS 19872 & LBG & VANDELS & 3.452 & NB396 & 9.6 & 5.6 & $-8.8$ & 111.8 \\ 
    CDFS 20745 & LBG & VANDELS & 3.495 & NB396 & 9.0 & 5.5 & $-8.3$ & 400.6 \\ 
    CDFS 24975 & LBG & VANDELS & 3.187 & NB3727 & 9.7 & 19.8 & $-8.5$ & 214.9 \\
    \hline
    \end{tabular}
    \label{tab:lyc-physical}
\end{table*}

\begin{table*}
    \centering
    \caption{Parameters required for the calculation of \fescrel\ and \fesc\ for candidate LyC leakers identified in this study.}
    \begin{tabular}{l c c c c c c c c}
    \hline
    CANDELS ID & $z$ & $(L_{1400}/L_{900})_{\rm{int}}$ & $A_V$ (mag) & $F_{900}$ ($\mu$Jy) &  $(F_{900}/F_{1400})_{\rm{obs}}$ & <exp($-\tau_{900}$)> & \fescrel & \fesc \\
    \hline
    CDFS 5161 & 3.420 & 3.7 & $0.6 \pm 0.1$ & $0.036 \pm 0.015$ & $0.07 \pm 0.02$ & 0.28 & $0.91 \pm 0.39$ & $0.53 \pm 0.24$ \\ 
    CDFS 9358 & 3.229 & 3.4 & $0.5 \pm 0.1$ & $0.048 \pm 0.020$ & $0.02 \pm 0.01$ & 0.32 &  $0.21 \pm 0.11$ & $0.14 \pm 0.07$ \\ 
    CDFS 9692 & 3.470 & 2.3 & $0.4 \pm 0.1$ & $0.039 \pm 0.020$ & $0.06 \pm 0.03$ & 0.27 & $0.55 \pm 0.25$ & $0.38 \pm 0.19$ \\ 
    CDFS 12448 & 3.460 & 4.1 & $0.1 \pm 0.1$ & $0.006 \pm 0.001$ & $0.02 \pm 0.01$ & 0.27 & $0.24 \pm 0.10$ & $0.20 \pm 0.12$ \\ 
    CDFS 13385 & 3.431 & 3.5 & $0.3 \pm 0.1$ & $0.009 \pm 0.002$ & $0.09 \pm 0.02$ & 0.28 & $1.12 \pm 0.24$ & $0.85 \pm 0.33$ \\ 
    CDFS 15718 & 3.439 & 2.0 & $0.4 \pm 0.1$ & $0.034 \pm 0.009$ & $0.06 \pm 0.02$ & 0.28 & $0.42 \pm 0.12$ & $0.28 \pm 0.10$ \\ 
    CDFS 16444 & 3.128 & 3.0 & $0.7 \pm 0.1$ & $0.068 \pm 0.014$ & $0.06 \pm 0.01$ & 0.34 & $0.55 \pm 0.11$ & $0.30 \pm 0.07$ \\ 
    CDFS 18454 & 3.163 & 2.2 & $0.8 \pm 0.1$ & $0.045 \pm 0.021$ & $0.04 \pm 0.02$ & 0.34 & $0.42 \pm 0.21$ & $0.20 \pm 0.10$ \\ 
    CDFS 19872 & 3.452 & 4.0 & $0.1 \pm 0.1$ & $0.040 \pm 0.020$ & $0.02 \pm 0.01$ & 0.28 & $0.32 \pm 0.16$ & $0.29 \pm 0.30$ \\ 
    CDFS 20745 & 3.495 & 4.0 & $0.2 \pm 0.1$ & $0.033 \pm 0.015$ & $0.03 \pm 0.02$ & 0.27 & $0.45 \pm 0.23$ & $0.38 \pm 0.30$ \\ 
    CDFS 24975 & 3.187 & 3.4 & $0.7 \pm 0.1$ & $0.043 \pm 0.021$ & $0.05 \pm 0.02$ & 0.33 & $0.53 \pm 0.21$ & $0.27 \pm 0.11$ \\ 
    \hline
    \end{tabular}
    \label{tab:lyc-fesc}
\end{table*}

It is important to note that given the stochasticity of the IGM attenuation at these redshifts \citep[see for example][]{bas21}, it is hard to accurately determine the absolute LyC \fesc\ for sources at $z\gtrsim3$. Therefore, the IGM stochasticity introduces an unavoidable element of scatter in the \fesc\ measurements, which in turn may be an important source of scatter in the correlations between \fesc\ and key galaxy properties that will be investigated in the sections that follow. We will discuss the effects of IGM stochasticity on the inferred LyC \fesc\ measurements in \S\ref{sec:discussion}.

Given the relatively large sample size of 183 sources, we expect $\approx4$ sources to appear as noise fluctuations at the $2\sigma$ level. From Table \ref{tab:lyc-fesc} we note that 4 out of 11 sources identified as candidate LyC leakers are close to the $2\sigma$ confidence level in their respective LyC images and could be consistent with noise fluctuations. Deeper photometric data at LyC wavelengths is needed to confirm the LyC leaking nature of these sources.
 
Apart from the 11 candidate LyC leakers, all other galaxies in our sample remain undetected at $\gtrsim2\sigma$ levels in the available LyC images, indicating that the global \fesc\ from star-forming galaxies at $z=3-3.5$ must be low. In the sections that follow we derive the global \fesc\ from different populations of galaxies in our sample through stacking the low-significance LyC fluxes measured from the majority of galaxies, and use these stacked measurements to set additional constraints on any global dependence of \fesc\ on key galaxy properties traced by sub-populations of galaxies within our sample.

\subsection{Stacked LyC \fesc\ measurements}
Only $\approx 6\%$ galaxies in our sample have a $\gtrsim2\sigma$ detection in the LyC images. Therefore, to constrain the global \fesc\, we must co-add the LyC signal from galaxies that were not classified as candidate LyC leakers. To do this, we first remove the 11 individual LyC leakers from the stacking sample and employ both the stacking of LyC images, as well as co-adding the LyC flux measured through aperture photometry as described below.

\begin{figure}
    \centering
    \includegraphics[width=0.48\textwidth]{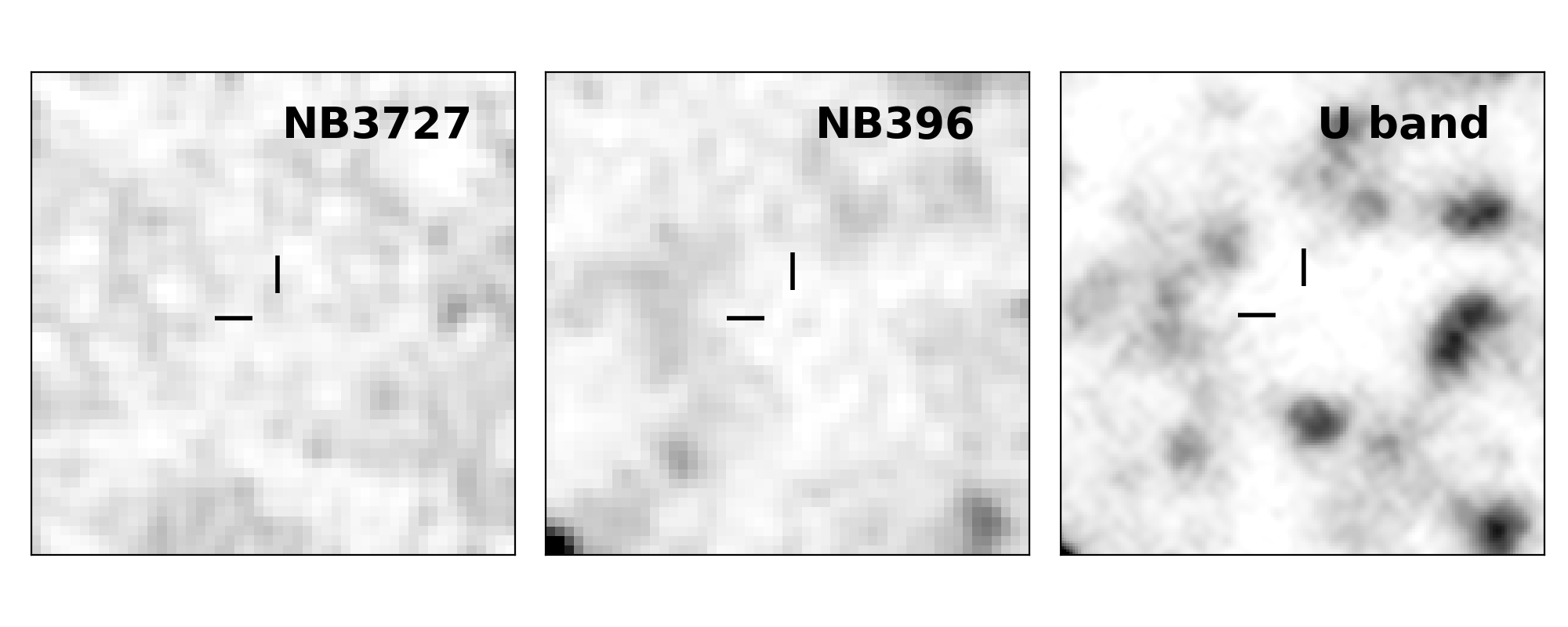}
    \vspace{-10pt}
    \caption{Stacked image from all sources in \emph{NB3727} (left), \emph{NB396} (middle) and \emph{U} band (right). We do not see any statistically significant flux in any of the stacked images.}
    \label{fig:stacks}
\end{figure}

We first investigate whether any LyC signal can be measured through stacking the LyC images of individual galaxies falling within the redshift coverage of each LyC filter. To do this, we produce $10''\times10''$ cutouts of both the LyC image and the corresponding root-mean-squared (RMS) error image centred on the CANDELS coordinates for galaxies that are not individual LyC leakers. We then apply $5\sigma$ clipping to the cutouts to masks bright pixels that could bias the stacked image. We then co-add the sigma-clipped cutouts using a weighted averaging method, where the corresponding RMS of each pixel of the galaxy cutout is used to assign a weight to the pixels of each galaxy cutout. In Figure \ref{fig:stacks} we show the results from stacking the LyC cutouts for each of the filters and as can be seen, we do not find any statistically significant LyC signal even in the stacked \emph{NB3727}, \emph{NB396} and \emph{U} band images, consistent with that reported by \citet{gua16} from the two narrow bands and by \citet{nai18} from the $U$ broad band. 

Instead of extracting a flux from these stacked LyC images and relying on the sample averaged values for the intrinsic non-ionising to ionising UV ratio, IGM transmissivity and dust attenuation to calculate \fesc, we use a slightly different approach and derive global \fesc\ using the standard deviation of the weighted average of the individual LyC fluxes measured through aperture photometry across the available LyC images as described in \S\ref{sec:photometry} \citep[see also][]{gua16}. The weight assigned to the \fesc\ measurement for each source is the inverse square of the error on \fesc, or $1/\Delta f_{\rm{esc}}^2$. 

After background subtraction, the weighted average of sources that are not classified as candidate LyC leakers would be expected to be close to zero, with the standard deviation providing a robust $1\sigma$ upper limit on the LyC signal from the sample \citep[see also][]{gua16}. This method takes full advantage of the availability of spectroscopic redshifts and accurate SED fitting that enable reliable measurements of both the values, errors and limits on individual \fescrel\ and \fesc\ for each galaxy. 

Galaxies in the redshift range covered by \textit{NB3727} do not overlap with the other two LyC filters. However, galaxies with redshifts between $3.41-3.49$ have overlapping LyC coverage in both \textit{NB396} and \textit{U} images. Therefore, to assign a LyC flux and \fesc\ to each unique galaxy in our sample, we prioritise the flux measured from the LyC image with a lower error, which translates to a lower error on \fesc. We note here that \fesc\ measurements/limits recorded using both LyC filters are broadly in agreement with each other. 

Using the standard deviation of the weighted average of LyC flux from the unique sample of galaxies that are not individually identified as LyC leakers, we obtain a $1\sigma$ upper limit of \fesc\,$<0.12$. We additionally calculate the stacked \fesc\ limit obtained from all galaxies in the sample (i.e. both LyC detections and non-detections) finding a slightly higher $1\sigma$ limit of \fesc\ $<0.16$, which is not surprising. This stacked measurement from the full sample potentially helps minimise the effects of stochastic IGM transmission along the line of sight, which we discuss further in \S\ref{sec:discussion}.

\subsubsection{LyC \fesc\ from LBG and blind selected galaxies}
Out of the 172 galaxies that were not identified as individual LyC detections, 103 are LBG selected and 69 are blindly selected. Using weighted averaging, we derive $1\sigma$ \fesc\ limits of $<0.11$ for the LBG selected and a slightly higher $1\sigma$ limit of $<0.13$ for the blind selected sample. We note here that since the blind selected sample is also fainter in UV magnitudes, the flux limited nature of the available LyC images may bias the inferred \fesc\ to higher values.

A LyC \fesc\,$<0.11$ for our LBG selected sample is comparable with \fesc\,$\approx0.09$ reported by \citet{ste18} by stacking deep UV spectra of $z\sim3$ LBGs and \fesc\,$\approx0.06$ obtained by a later reanalysis of the \citet{ste18} sample by \citet{pah21}, and with other measurements from LBGs at $z\sim3$ \citep[e.g.][]{gra16, gua16, mar17, nai18, mes21}. The slightly higher $1\sigma$ \fesc\ limit of \fesc\,$<0.13$ measured for blind-selected galaxies is also consistent with measurements of emission-line selected galaxies at $z=2-4$ \citep[e.g.][]{mos15, jap17, ji20, smi20}.

\subsubsection{LyC \fesc\ from extreme emission line galaxies (EELGs)}
\label{sec:eelgs}
In this section we investigate the stacked \fesc\ measurements of the previously identified extreme emission line galaxies (EELGs; EW$_0$(\ohb) $\ge300$\,\AA) as well as normal emission line galaxies (ELGs; EW$_0$(\ohb) $<300$\,\AA) in our sample (see \S\ref{sec:eelg-id}). For this analysis we include the 11 candidate LyC leakers to better understand the dependence of \fesc\ on \ohb\ EWs.

There are 59 EELGs and 124 ELGs in our sample. The average EW$_0$(\ohb) for EELGs is $421.2$\,\AA\ and the average EW$_0$(\ohb) for ELGs is $183.5$\,\AA. By once again calculating the weighted average of \fesc\ for ELGs and EELGs, we determine $1\sigma$ upper limits of \fesc $<0.14$ for ELGs and \fesc $<0.20$ for EELGs. We note here that we obtain slightly higher \fesc\ limits from bins created based on emission line strengths compared to the LBG and blind-selected samples because the individually identified LyC leakers are included in the ELG and EELG bins.

Qualitatively speaking, the distribution of \ohb\ EWs measured for individual LyC leakers is similar to the distribution of the parent sample, implying that the individual LyC leakers seem to have been randomly drawn from the parent sample at least in terms of \ohb\ EWs. This is consistent with what was reported by \citet{nak20} for a sample of LAEs at $z\sim3$, where the \ohb\ EWs of LAEs that are LyC leakers and non-leakers are not significantly different.

In Table \ref{tab:subsets} we give the median galaxy physical properties and the measured \fesc\ for subsets created based on the initial spectroscopic selection method, as well as on the strength of \ohb\ emission lines. No clear differences in \fesc\ are seen between these subsets, posing questions surrounding the origin of leaking LyC radiation from only a low fraction of galaxies at intermediate redshifts. In the following section we investigate the dependence of \fesc\ on key galaxy properties for both candidate LyC leakers and stacks of galaxies.
\begin{table*}
    \centering
    \caption{Physical properties and \fesc\ for LBG or blind selected galaxies, ELGs and EELGs, and the full sample. Note that the LBG and blind selected subsets are created only for those galaxies that are undetected in the LyC images at $\gtrsim2\sigma$ level, but the ELG and EELG subsets include the individual LyC leakers as well.}
    \begin{tabular}{l c c c c c}
    \hline 
    Subset & $N$ & log($M_\star$/$M_\odot$) & log(sSFR/yr$^{-1}$) & EW$_0$ (\ohb) (\AA) & \fesc \\    \hline
    LBG selected, LyC non-detections & 103 & $9.4$ & $-8.2$ & $236.2$ & $<0.11$ \\
    Blind selected, LyC non-detections & 69 & $9.0$ & $-8.3$ & $207.2$ & $<0.13$ \\ 

    ELGs (all) -- EW$_0$(\ohb) $<300$\,\AA & 124 & $9.3$ & $-8.3$ & 183.5 & $<0.14$ \\
    EELGs (all) -- EW$_0$(\ohb) $\ge300$\,\AA & 59 & $8.9$ & $-8.0$ & 421.2 & $<0.20$ \\

    Full sample & 183 & $9.2$ & $-8.3$ & 230.9 & $<0.16$ \\
    \hline
    \end{tabular}
    \label{tab:subsets}
\end{table*}

\section{Results}
\label{sec:results}
Having measured LyC fluxes and \fesc\ values for every source in the sample, and having used stacking to establish \fesc\ limits from the sample of galaxies that do not show any LyC leakage, in this section we explore potential correlations between \fesc\ and the properties of the underlying stellar population in the galaxies. Constraining the dependence of \fesc\ on these key galaxy properties is important to understand the key drivers of reionisation at $z>6$, and here we rely on robust SED-fitting based measurements to explore these.

\subsection{Dependence of \fesc\ on stellar mass}
\begin{figure}
    \centering
    \includegraphics[scale=0.4]{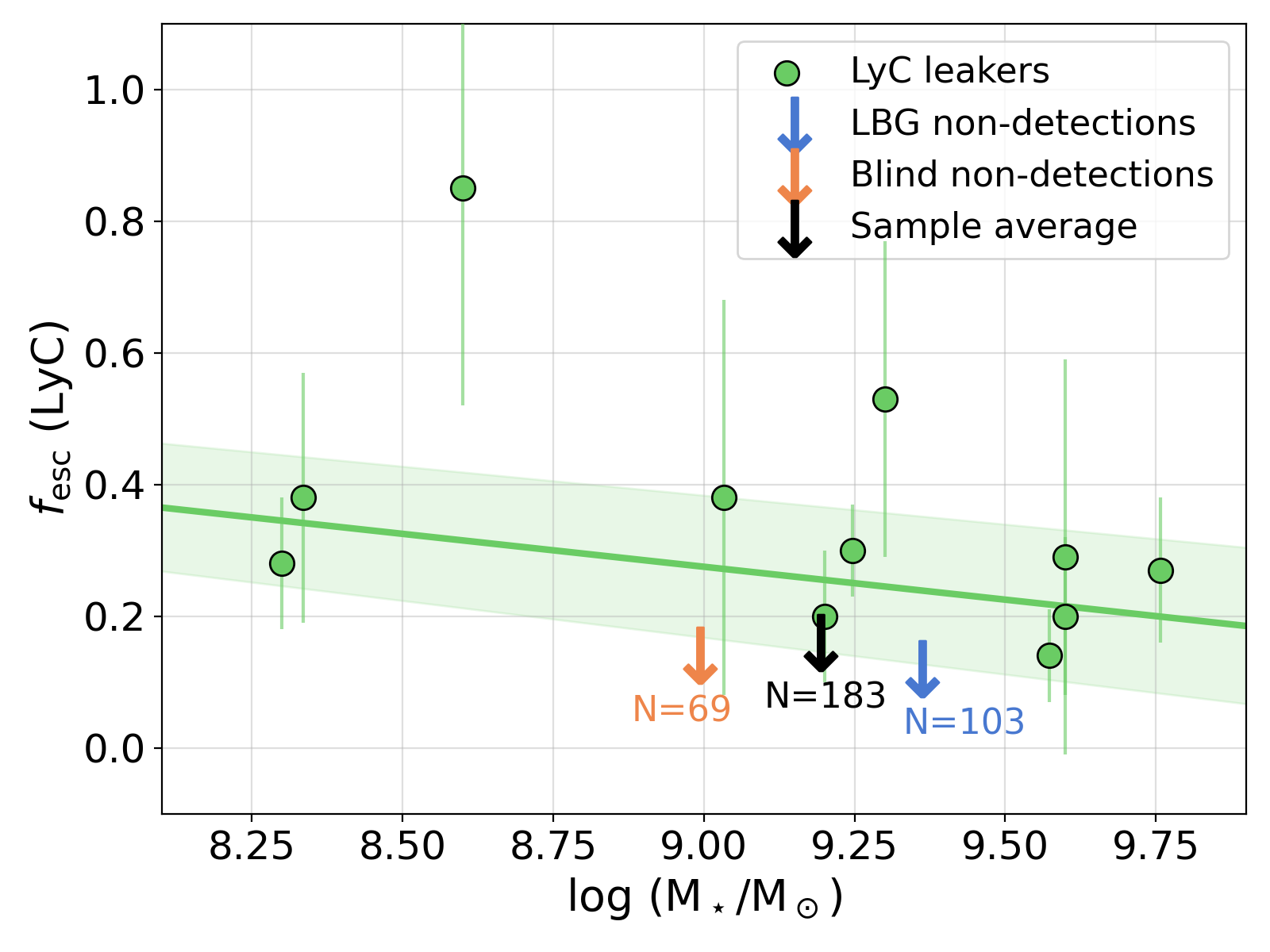}
    \caption{LyC \fesc\ as a function of stellar mass for candidate LyC leakers in our sample (green points), and stacked measurements from LBG selected (blue arrow), blind-selected galaxies (orange arrow) and the sample average (black arrow) in this study. The number of galaxies used in the stacking measurement are also given. We find a very weak correlation between \fesc\ and stellar mass, with a best-fitting slope $-0.10\pm0.02$ (solid green line, uncertainty in shaded region). This dependence on stellar mass is stronger that the recently observed lack of correlation between the two quantities for a sample of low mass ($<10^8\,M_\star$) galaxies at $0.3 \lesssim z \lesssim 0.45$ by \citet{izo21}, however the scatter is very large. We note that the stacked limits are not included when obtaining the best-fitting relations in this study.}
    \label{fig:lyc-stellarmass}
\end{figure}

We first explore the dependence of \fesc\ on the stellar masses of galaxies. In Figure \ref{fig:lyc-stellarmass} we show the \fesc\ measured for the candidate LyC leaking candidates as well as the stacked limits obtained for the sub samples of LBG and blind selected galaxies in our sample. As reported in \S\ref{sec:sed-fitting}, the median stellar mass of LBG selected galaxies is $10^{9.3}\,M_\odot$ and that of the blind selected galaxies is $10^{9.0}\,M_\odot$.  

We find that the candidate LyC leaking galaxies occupy a large range of stellar masses, ranging from $10^{8.3} - 10^{9.7}\,M_\odot$. The galaxy with the highest \fesc\ in our sample, CDFS 13385 at $z=3.431$ with \fesc\ $=0.85\pm0.33$ has a stellar mass of $\approx10^{8.3}\,M_\odot$, which is lower than the stellar masses of known LyC leaking galaxies with \fesc $>0.5$ at $z\gtrsim3$ such as \emph{Ion2} \citep{deb16, van16a}, \emph{Ion3} \citep{van18} and Q1549$-$C25 \citep{sha16}, and comparable to the stellar masses of LyC leaking LAEs identified by \citet{fle19}. The LyC leaker with the highest stellar masses in our sample has $M_\star \approx 10^{9.7}$ -- CDFS 24975 at $z=3.187$ with \fesc\ $=0.27\pm0.11$. 

We use least-squares fitting to obtain the best-fitting parameters for the function \fesc\ $= A \log(M_\star/M_\odot) + B$, finding $A = -0.10 \pm 0.02$ and $B = 1.17 \pm 0.40$ (solid green line in Figure \ref{fig:lyc-stellarmass}). The slope of the best-fitting relation suggests a very weak dependence of \fesc\ on the stellar mass of galaxies in our sample. Although not very strong, the mass dependence of \fesc\ that we report from our candidate LyC leakers is stronger than the no apparent mass dependence reported by \citet{izo21} for a sample of very low mass ($M_\star < 10^8$\,$M_\odot$) galaxies in the redshift range $z=0.31 - 0.45$.

We also find that there is no clear difference in the upper limits of \fesc\ obtained by stacking samples of LBG-selected and blind-selected galaxies. Blind-selected galaxies on average have lower stellar masses, and a slightly higher \fesc\ upper limit, which is consistent with the mass dependence obtained by fitting measurements for the candidate LyC leakers in our sample.

\subsection{Dependence of \fesc\ on specific SFR}
\begin{figure}
    \centering
    \includegraphics[scale=0.4]{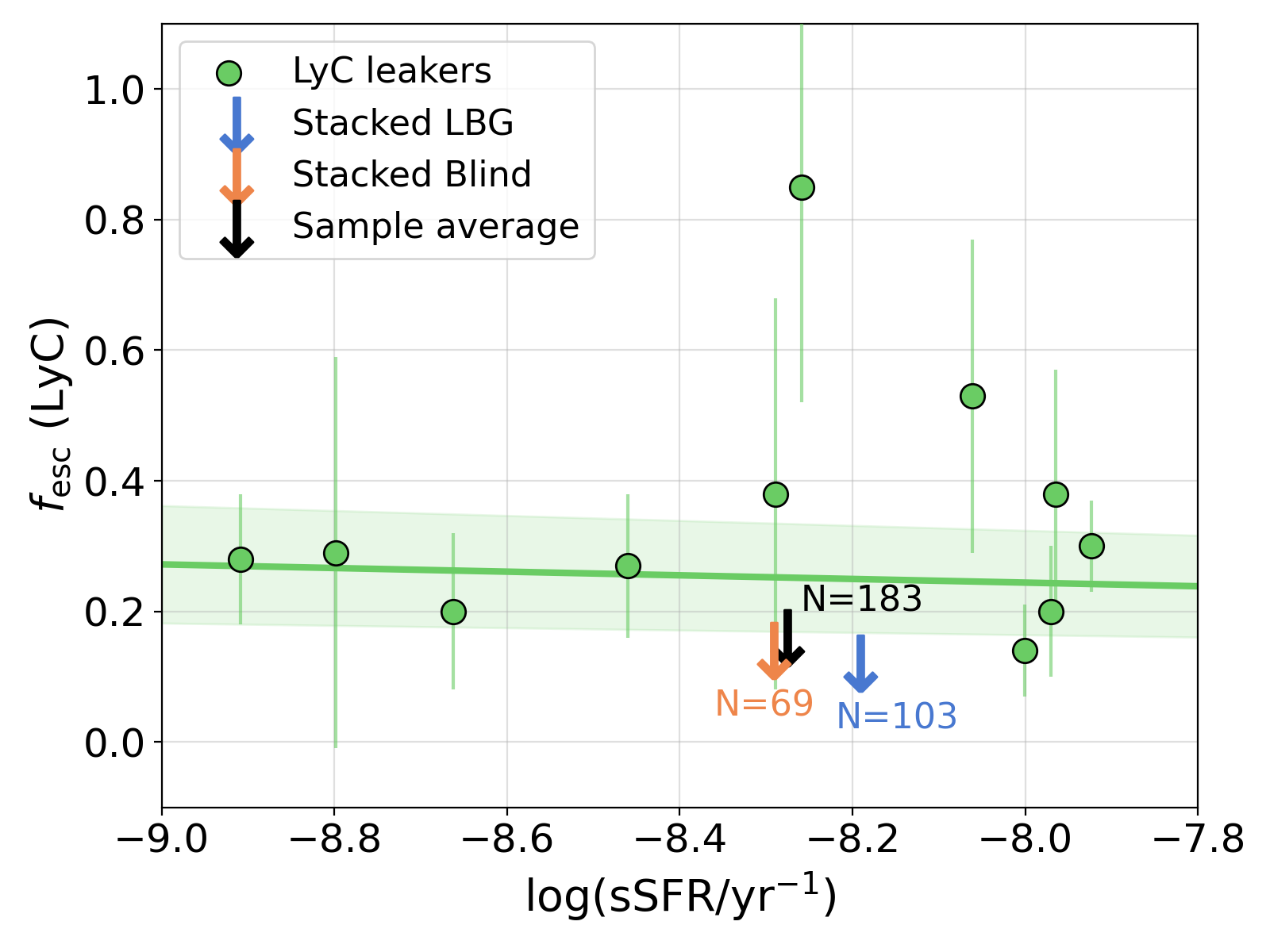}
    \caption{Distribution of \fesc\ and sSFR of candidate LyC leakers, LBG and blind-selected galaxies as well as the sample average (same legend as Figure \ref{fig:lyc-stellarmass}). We find little to no correlation between \fesc\ and sSFR of individual LyC leakers at the $2\sigma$ confidence level, with a best-fitting slope $-0.028\pm0.015$ (solid green line with uncertainty in shaded region). The stacked \fesc\ values derived from LBG and blind-selected samples are consistent with the best-fitting relation. The lack of dependence of \fesc\ on sSFR has also been seen in simulations \citep[e.g.][]{yoo20}.}
    \label{fig:lyc-ssfr}
\end{figure}

We now investigate correlations between \fesc\ and specific SFR (sSFR), which is the star-formation rate per unit stellar mass of a galaxy. The distribution of \fesc\ and sSFR for both the candidate LyC leakers and stacks of LBG and blind selected galaxies is shown in Figure \ref{fig:lyc-ssfr}. As reported in \S\ref{sec:sed-fitting}, the blind selected galaxies in our sample have a slightly lower median specific SFR of log(sSFR/yr$^{-1}$) $= -8.3$ compared to log(sSFR/yr$^{-1}$) $= -8.2$ for LBG selected galaxies.

The candidate LyC leakers cover a large range of sSFRs, from log(sSFR/yr$^{-1}$) $\approx -8.8$ to log(sSFR/yr$^{-1}$) $\approx -7.9$. We note that the majority of LyC leaking candidates (seven out of 11) have relatively high sSFR values, with log(sSFR/yr$^{-1}$) $\gtrsim -8.1$, comparable to LyC leaking LAEs from \citet{fle19}. In particular, the strongest LyC leaker in our sample, CDFS 13385, has log(sSFR/yr$^{-1}$) $= -8.1$, which is lower than \emph{Ion2} and \emph{Ion3} but more than an order of magnitude higher than Q1549$-$C25 \citep{sha16}.

To quantify the dependence of \fesc\ on sSFR, we again use least-squares fitting and obtain best-fitting correlations to the function \fesc\ $= C \log(\rm{sSFR}/\rm{yr}^{-1}) + D$ with a slope of $C=-0.028\pm 0.015$ (solid green line in Figure \ref{fig:lyc-ssfr}). This best-fitting suggests a very weak decrease of \fesc\ with increasing sSFR, but is consistent with no dependence of \fesc\ on sSFR at the $\approx2\sigma$ confidence level, consistent with what has been reported from disc galaxy simulations at high redshifts \citep[e.g.][]{yoo20}.

Assuming that the stellar mass of a galaxy is closely linked to its size at UV wavelengths, no strong dependence of \fesc\ on sSFR may also translate to no strong dependence of \fesc\ on the star-formation surface density of galaxies \citep[e.g. Model II of][]{nai20}. This may have bearing on the dominant contribution of the UV brightest sources towards reionisation, and accurate size measurements of our candidates LyC leakers that will be explored in a future study will help robustly test the dependence of \fesc\ on SFR surface densities.

\subsection{Dependence of \fesc\ on \ohb\ strengths} 
\begin{figure}
    \centering
    \includegraphics[scale=0.40]{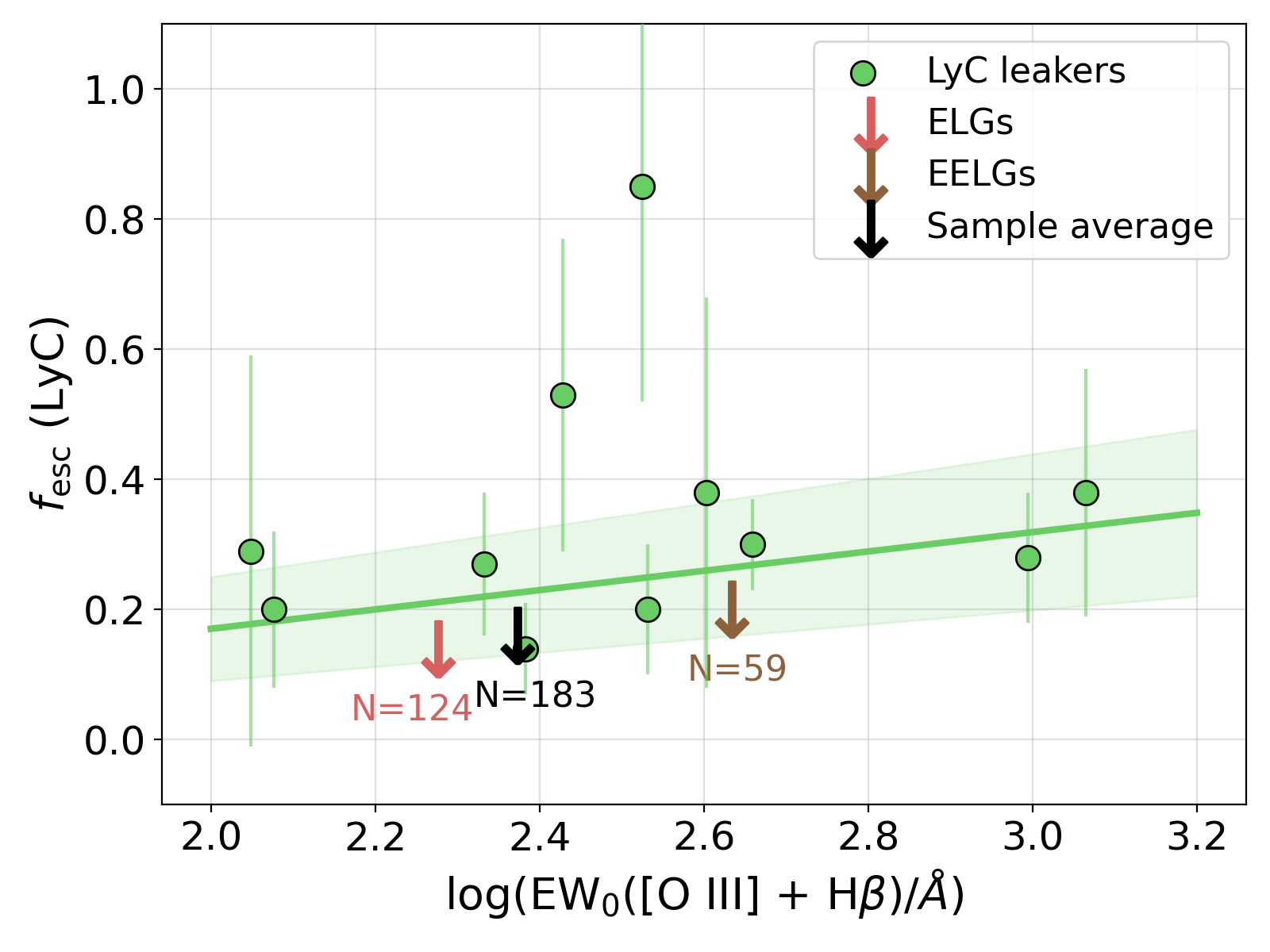}
    \caption{Distribution of \fesc\ and log(EW$_0$(\ohb)/\AA) for individual sources (green points) and limits derived for ELGs (red arrow) and EELGs (brown arrow) in our sample. We find a positive correlation between \fesc\ and the strength of the \ohb\ lines for the candidate LyC leaker candidates, with best-fitting slope of $0.15 \pm 0.04$. High \ohb\ line fluxes are seen when the physical conditions within a galaxy favour low metallicities or intense star-formation activity, which are important conditions to both produce a large number of LyC photons as well as facilitate a high escape fraction of these photons.}
    \label{fig:lyc-o32}
\end{figure}

In this section we investigate the dependence of \fesc\ on the \ohb\ line strengths for galaxies in our sample. In Figure \ref{fig:lyc-o32} we show the \fesc\ and the log of EW$_0$(\ohb) for individual LyC leakers (green points) as well limits derived for normal emission line galaxies (ELGs; red arrow) and and extreme emission line galaxies (EELGs; brown arrow), which were defined in \S\ref{sec:eelgs}. We note that 6 out of the 11 individual LyC leakers can be classified as EELGs, with EW$_0$(\ohb) $>300$\,\AA.

For the individual LyC leaking candidates, CDFS 9692 at $z=3.470$ with \fesc\ $=0.38\pm0.19$ has the highest EW$_0$(\ohb) value of $1160.3$\,\AA. The strongest LyC leaker, CDFS 13385, also shows extreme \ohb\ line fluxes, with EW$_0$(\ohb) $=986.8$\,\AA, in line with expectations from galaxies with high \fesc. However, overall we do not find consistently high \ohb\ EWs across our candidate LyC leaking galaxies. Similarly, as reported in \S\ref{sec:eelgs} we do not find a considerably higher \fesc\ using stacking for EELGs compared to the sample of ELGs, but do find that the \fesc\ limit for EELGs is higher than that for ELGs.

A best-fitting relation between \fesc\ and \ohb\ line strengths shows a positive correlation, indicating that galaxies with higher \fesc\ also tend to show stronger \ohb\ lines, shown in Figure \ref{fig:lyc-o32}. The best-fitting relation has a positive slope of $0.15\pm0.04$, indicative of a mild dependence of \fesc\ on rest-frame \ohb\ EW, with the general trend of increasing \fesc\ at higher \ohb\ strengths consistent with what has been seen in the local Universe \citep[e.g.][]{izo18a}. The higher \fesc\ upper limit of $<0.20$ seen for EELGs compared to \fesc\ $<0.14$ seem for ELGs is also consistent with the best-fitting relation obtained.

NIR spectroscopy of LyC leaking LAEs from the \citet{fle19} sample presented by \citet{nak20}, however, did not find uniformly high \ohb\ line strengths for LyC leakers, which is consistent with our findings. \citet{izo18a} reported a large spread in the \fesc\ values of local LyC leaking galaxies with high \oiii\ line strengths. Therefore, there is increasing evidence that suggests that high \ohb\ line strengths may be a necessary condition for \fesc, but are by no means sufficient to guarantee high \fesc. This is reflected in our galaxy sample as well, where galaxies with extremely high \ohb\ strengths do not all necessarily appear to be strong LyC leakers.

We also note here that given the volume limited nature of the spectroscopic galaxy survey data used in this study, and the rarity on the sky of extreme line emitting galaxies, we are likely to miss several extreme \ohb\ line emitters in our sample. Studies investigating correlations between \fesc\ and strength of \ohb\ emission (or \oiii/\oii\ ratios) have largely focused on galaxy samples at relatively low redshifts, and the lack of completeness of extreme line emitters that are likely to exhibit high \fesc\ is a caveat of any attempt to investigate the dependence of \fesc\ on \ohb\ strength at $z>3$ from our data set.


Therefore, our analysis has found a weak negative correlation between \fesc\ and stellar mass, but contrary to expectations of models of reionisation, we do not see any significant correlation between \fesc\ and specific SFR of galaxies. We have also found that only a small fraction ($\approx6\%$) of galaxies in our sample can be reliably classified as candidate LyC leakers, although the uncertainties on the measured \fesc\ are large owing to low significance of the measured LyC flux. The question that naturally arises following these findings, and is paramount to understanding the nature of sources in the early Universe that drove the bulk of reionisation, is: What regulates \fesc\ in a galaxy? We discuss the relevance and implications of our findings in the following section.

\section{Discussion -- What regulates \fesc?}
\label{sec:discussion}
The escape fraction of ionising photons from star-forming galaxies crucially depends on the fraction of ionising radiation absorbed/scattered by the neutral gas and dust around the photon emitting regions. To have high \fesc, a large amount of neutral gas and dust must have been expelled from the star-forming regions through either supernova or AGN feedback \citep[e.g.][]{wis14, tre17, tre18, bar20, kat20}. Given the dependence of high \fesc\ from star-forming galaxies on the presence of ionised channels through which LyC photons can escape into the IGM, the orientation of the galaxies and potential (mis)alignment of these channels with the observer will play a key role in whether or not a galaxy appears to be leaking LyC radiation \citep[e.g.][]{bor14}. 

Therefore, by selection only a fraction of intermediate redshifts reionisation era analogue galaxies in any given sample will be identified as LyC leakers, assuming that these galaxies uniformly leak non-zero fractions of LyC photons. This geometrical misalignment of the viewing angles of a majority of sources may explain the relatively low numbers of individual galaxies identified in our sample, and across other samples that appear to be leaking significant fractions of LyC photons \citep[e.g.][]{fle19}. A further explanation to the relatively low \fesc\ values seen in the more relatively massive star-forming galaxies at intermediate redshifts could be the higher metallicities compared to the ultra-low metallicities expected from galaxies in the reionisation era \citep{yoo20}.

Another important factor that has an impact on whether or not a galaxy selected from a large sample appears to be a LyC leaker is the stochasticity of the IGM along the line-of-sight. Using ensembles of simulated IGM transmission functions along multiple lines-of-sight at $z\gtrsim3$, \citet{bas21} showed that for a given underlying distribution of LyC \fesc\ in a galaxy sample, only those galaxies that lie along sightlines with higher IGM transmission appear to be leaking LyC radiation, exhibiting high \fesc. Inclusion of a possible IGM transmission bias for both candidate LyC leakers and non leakers in this work may result in reduced \fesc\ values for the leakers, with an increase in the upper limits of \fesc\ inferred from stacking the non-leakers \citep[e.g.][]{bas21}.

In addition to observational effects such as orientation or stochastic IGM transmission, theoretical modelling has also revealed complex relationships between stellar population properties such as metallicity, ionisation parameter, etc. and \fesc, which impact the observed \fesc\ and its dependence on physical properties of stars in a star-forming galaxy \citep[e.g.][]{bas19}. Further complexities are introduced by geometrical and viewing angle effects that further impact correlations between \fesc\ and the strength of the ionising radiation field and ionising photon production rates from young stars traced by strong \ohb\ line strengths \citep[e.g.][]{kat20}, among other galaxy properties. \citet{bar20} showed that the periods during which a galaxy exhibits intense star-formation activity and ionising photon production, and periods corresponding to high \fesc\ may be non-coincident, with high periods of \fesc\ often following starburst activity \citep[see also][]{kim14, tre17, kak21}. 

However, it is important to note that conditions enabling high \fesc\ following a starburst event may not necessarily coincide with an equally high production rate or efficiency of ionising photons ($\xi_{\rm{ion}}$), as also seen in the simulations of \citet{bar20}. Therefore, there is an element of fortuitousness introduced in whether or not a star-forming galaxy selected from a large sample at any given epoch appears to be leaking LyC photons, and crucially depends on periods of both high $\xi_{\rm{ion}}$ as well as high \fesc\ being coincident \citep[see also][]{fle19, nak20}.

Further, a spatially resolved analysis of the lensed LyC leaker dubbed `Sunburst' at $z=2.37$ \citep{dah16, riv19} recently showed that a young massive star cluster dominates the leaking LyC radiation from the entire galaxy (\fescrel\,$\approx0.43-0.93$), whereas older clusters show considerably lower LyC \fesc\ values \citep{van21}. Without magnification due to lensing, the spatially unresolved Sunburst arc would exhibit a much lower LyC leakage overall (\fescrel\,$\approx0.06-0.12$) as the LyC radiation escaping along other lines of sight would be missed due to insufficient LyC imaging depths \citep[see also][]{van20b}. 

Therefore, in a scenario where a large fraction of LyC photons produced by only a limited number of individual young star clusters within galaxies manage to escape into the IGM through favourably aligned channels, there would be increased scatter in the observed dependence of \fesc\ and galaxy properties such as stellar mass or specific SFR for any given sample of galaxies. Although this would explain the lack of significant correlations observed between \fesc\ and the physical properties of candidate LyC leakers in this study, most importantly with the sSFR, such an effect complicates the observational understanding of what regulates the escape of ionising photons and consequently, which sources were the key drivers of reionisation. This makes studying indicators that trace the presence of LyC photons escape channels and their dependence on key galaxy properties crucial to understanding the physical and geometrical conditions that enable a high apparent \fesc\ in the epoch of reionisation \citep[e.g.][]{ste18}. 

One such probe is the \lya\ line, whose shape, width and intensity can offer insights into LyC and \lya\ escape channels from clusters of young stars within high redshift galaxies \citep[e.g.][]{van18, van20b, ver17, izo18a, izo18b, mat18, mey21, nai21}. Additional high ionisation emission lines like \heii\,$\lambda1640$, \civ\,$\lambda\lambda1548,1550$ and \ciii\,$\lambda\lambda1907,1909$ in the rest-UV may help trace stellar populations that produce copious amounts of ionising photons, possibly leading to higher \fesc\ \citep[e.g.][]{sch18, ber19, sen19, lle21}. \citet{sax20} investigated \heii\ emitting star-forming galaxies in the redshift range $z\sim2.5-5$ and did not find any strong dependence of the presence or the strength of \heii\ emission on the stellar mass or star-formation rates of galaxies \citep[see also][]{nan19}, which may be consistent with factors that lead to a lack of observed dependence of \fesc\ on these properties as discussed above. 

A key advantage of our galaxy sample at $z\approx3-3.5$ is the availability of rest-frame UV spectra from various spectroscopic surveys, crucially also covering the \lya\ emission line, and in a future study we will explore the correlations between \fesc, \lya\ and other rest-UV spectroscopic indicators for these newly discovered candidate LyC leakers (and non-leakers). \emph{JWST} will give access, for the first time, to the full suite of rest-frame optical emission lines in $z>3$ galaxies, enabling direct measurements of the ionising radiation field strengths produced by young stars (through \oiii\ line strengths and \oiii/\oii\ ratios) as well as measurements of the ionising photon production efficiencies using H$\alpha$ measurements, which will considerably aid in identifying physical conditions within star-forming galaxies that lend themselves to high \fesc. The correlations between \fesc\ and rest-frame optical emission established at these intermediate redshifts will be instrumental in interpreting the ionising photon escape of galaxies at $z=6-9$, for which \textit{JWST} will enable rest-optical spectroscopy, in a bid to understand the nature of the main drivers of reionisation.

Therefore, going forward it is essential to obtain deeper LyC imaging for intermediate redshift analogues of reionisation era galaxies to properly account for the viewing angle and orientation effects when measuring \fesc. Further, detailed studies of the rest-frame UV and optical spectra of known LyC leakers are needed to investigate the presence of LyC escape channels and study the dependence of these features on galaxy properties. Better insights into the timescales and the incidence of high \fesc\ within star-forming galaxies may be needed from high-resolution simulations \citep[e.g.][]{mau21} of a larger number of reionisation era galaxies (with sufficient time resolution) to temporally tie periods of intense star-formation to \fesc\ and other galaxy properties.

\section{Summary and conclusions}
\label{sec:conclusions}
We have measured the Lyman continuum (LyC) escape fraction (\fesc) using ground based LyC images available for 183 spectroscopically confirmed galaxies in the CDFS/GOODS-S field in the redshift range $3.11<z<3.53$. The spectroscopic redshifts are taken from publicly available databases, that primarily targeted the higher mass Lyman break (LBG) selected galaxies or blindly discovered lower mass galaxies thanks to emission line surveys. 

We use available deep \textit{HST} photometry and colours at non-ionising UV wavelengths to remove possible foreground contaminants that may strongly bias the inferred LyC flux and \fesc\ as well as accurately constrain the spectral energy distributions (SEDs) of our galaxies giving robust measurements of key galaxy properties. Using a combination of stellar and nebular emission, we also derive rest-frame optical emission line strengths, primarily the \oiii\ and \hb\ line fluxes (that fall in the observed $K$ band) for our galaxies. 

We then measure LyC flux and \fesc\ using aperture photometry and detect LyC flux at $\ge 2\sigma$ significance levels in 11 out of 183 galaxies, with \fesc\ values ranging from $0.14-0.85$. Around $94\%$ of the galaxies in our sample remain undetected at the $\gtrsim\sigma$ level, and using weighted averaging of their LyC \fesc\ limits, we place a $1\sigma$ upper limit of \fesc\,$<0.12$ from the galaxies with no significant detections in the available LyC images. We additionally compare the weighted average \fesc\ limits from sub-samples of galaxies that were LBG or blind-selected, finding $1\sigma$ limits of \fesc\,$<0.11$ for LBG-selected and \fesc\,$<0.13$ for blind-selected galaxies. We further place \fesc\ limits from normal emission line galaxies (ELGs; \ohb\,$<300$\,\AA) and extreme emission line galaxies (EELGs; \ohb\,$>300$\,\AA), finding $1\sigma$ limits of \fesc\ $<0.14$ for normal and \fesc\ $<0.20$ for extreme emission line galaxies. 

Although the uncertainties on the measured \fesc\ values for the 11 candidate LyC leaking galaxies are high and a handful of sources have low signal-to-noise ratios in the LyC images ($\approx2\sigma)$, we find a weak negative correlation between \fesc\ and stellar mass at the $2\sigma$ level, with a shallow slope of $-0.10 \pm 0.02$. We do not find any significant correlation between \fesc\ and specific star-formation rate of the individual LyC leakers at the $2\sigma$ level, with a slope of $-0.028 \pm 0.015$. Lastly, we find a positive correlation between \fesc\ and EW$_0$(\ohb) for the candidate LyC leakers in our sample, suggesting that the \ohb\ line strengths may be important indicators of \fesc\ as has been previously reported in the literature. 

From our results, it remains unclear what the dominant galaxy property is when it comes to  regulating \fesc\ in LyC leaking galaxies. The low detection fraction of LyC leakers across our sample is consistent with a scenario where LyC photons escape through discrete ionised channels within star-forming galaxies, which are seldom oriented along the line-of-sight of the observer. The time-dependent nature of high \fesc\ that often follows periods of intense star-formation also suggests that only a low fraction of galaxies at any given epoch would be actively leaking LyC photons, consistent with our observations. 

The stochasticity of the intervening neutral intergalactic medium, which may play an important role in absorbing leaking LyC photons along the line-of-sight is another factor that may impact LyC detections. Future work tying LyC escape to the presence of ionised channels within galaxies using observations, exploration of the time-dependent nature of LyC leakage from larger samples of simulated reionisation era galaxies as well as investigation of IGM stochasticity across redshifts are needed to fully understand these effects in the context of contribution of star-forming galaxies towards the cosmic reionisation photon budget.

In conclusion, our work has discovered a sample of 11 previously undiscovered candidate LyC leakers at $z\sim3-3.5$ adding crucial statistics that are needed to understand the escape of ionising photons from intermediate redshift analogues of $z>6$ galaxies. Studying lower redshift analogues of reionisation era galaxies, where direct detection of escaping LyC radiation is possible, remains the only way to constrain LyC escape from star-forming galaxies that are expected to drive the bulk of reionisation. An increased sample of LyC leaking galaxies complemented by deeper LyC/UV imaging, higher resolution spectra across wavelengths and zoom-in simulations will help understand the unique physical conditions that drive high \fesc, bringing us a step closer towards unravelling the nature of galaxies and their ionising radiation that drove cosmic reionisation.

\section*{Acknowledgements}
We thank the referee for constructive feedback that improved the quality of this work. AS thanks Ross McLure and Mario Nonino for useful input and discussions, Fergus Cullen for sharing NIRVANDELS measurements and Mengtao Tang for sharing the best-fitting relation for emission line strengths from their analysis. AS and LP thank Alice Shapley for useful comments and discussions that improved the analysis presented. 

AS and RSE acknowledge financial support from European Research Council Advanced Grant FP7/669253. DS and ASL acknowledge support from Swiss National Science Foundation. RA acknowledges support from ANID FONDECYT Regular 1202007. MLl acknowledges support from the National Agency for Research and Development (ANID)/Scholarship Program/Doctorado Nacional/2019-21191036.

This work makes use of VANDELS data. The VANDELS collaboration acknowledges the invaluable role played by ESO staff for successfully carrying out the survey. This work is based on VUDS data, obtained with the European Southern Observatory Very Large Telescope, Paranal, Chile, under Large Program 185.A-0791, and made available by the VUDS team at the CESAM data center, Laboratoire d'Astrophysique de Marseille, France. This work is based on observations taken by the MUSE-Wide and MUSE-HUDF Surveys as part of the MUSE Consortium. This work is based on observations taken by the 3D-HST Treasury Program (GO 12177 and 12328) with the NASA/ESA HST, which is operated by the Association of Universities for Research in Astronomy, Inc., under NASA contract NAS5-26555. MLl acknowledges support from the National Agency for Research and Development (ANID)/Scholarship Program/Doctorado Nacional/2019-21191036.

This work has made extensive use of Jupyter and IPython \citep{ipython} notebooks, Astropy \citep{astropy, astropy2}, matplotlib \citep{plt}, seaborn \citep{seaborn}, pandas \citep{pandas} and \textsc{topcat} \citep{topcat}. This work would not have been possible without the countless hours put in by members of the open-source developing community all around the world, with special thanks to the Stackoverflow community.

\section*{Data availability}
The VANDELS data can be accessed using the VANDELS database at \url{http://vandels.inaf.it/dr4.html}, or through the ESO archives. The VUDS data can be accessed from the VUDS webpage: \url{http://cesam.lam.fr/vuds/} or ESO archives. MUSE data can be accessed from \url{https://musewide.aip.de/project/} and \url{http://muse-vlt.eu/science/udf/} and the 3D-HST data can be accessed from \url{https://archive.stsci.edu/prepds/3d-hst/}. The VIMOS10 data are available from \citet{bal10}. All the code used to perform the analysis in this paper was written in \textsc{python} and will be shared upon reasonable request to the corresponding author.




\bibliographystyle{mnras}
\bibliography{LyC} 


\appendix
\section{LyC and HST Cutout galleries with best-fitting SEDs of candidate LyC leakers}
\label{appendix:a}

\begin{figure}
    \centering
    \includegraphics[width=0.48\textwidth]{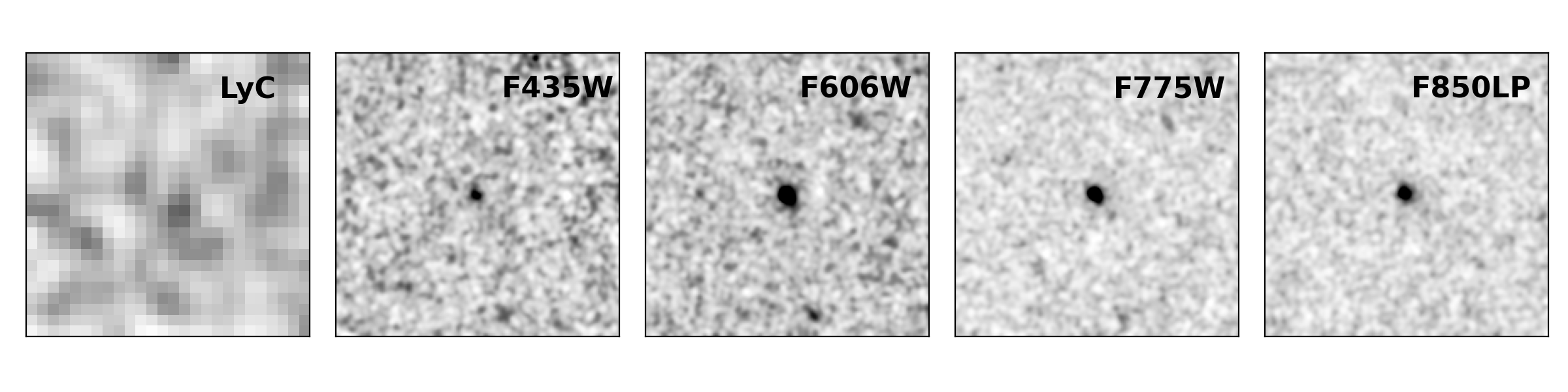}

    \vspace{-8pt}
    \includegraphics[width=0.48\textwidth]{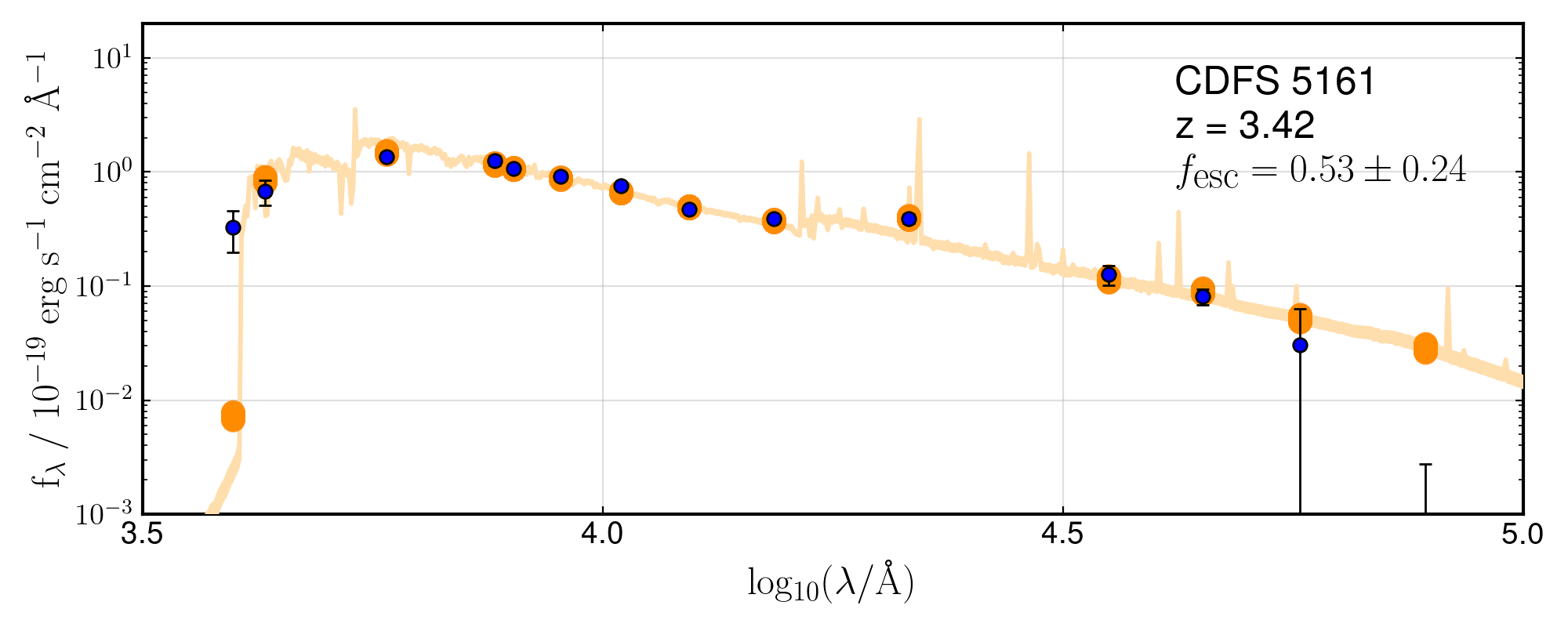}

    \includegraphics[width=0.48\textwidth]{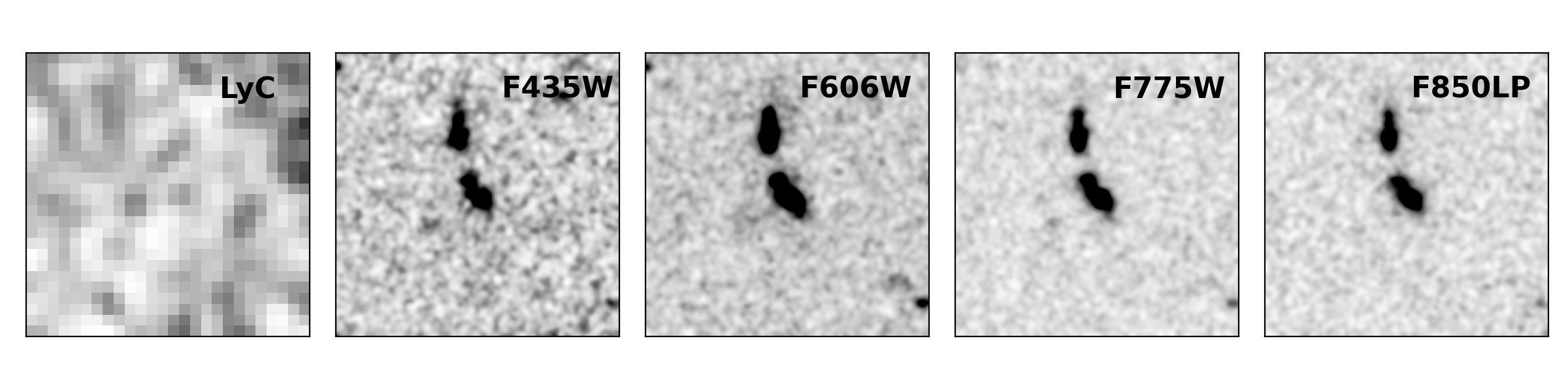}

    \vspace{-8pt}
    \includegraphics[width=0.48\textwidth]{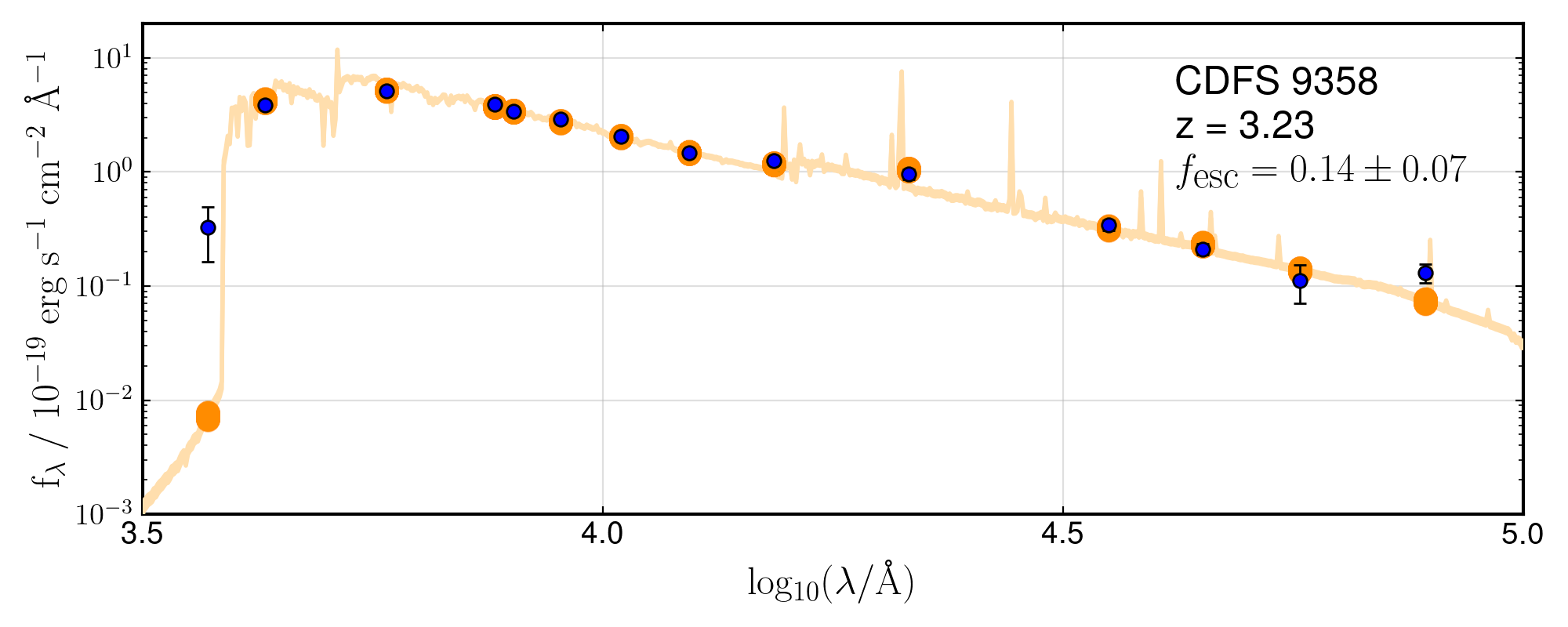}

    \includegraphics[width=0.48\textwidth]{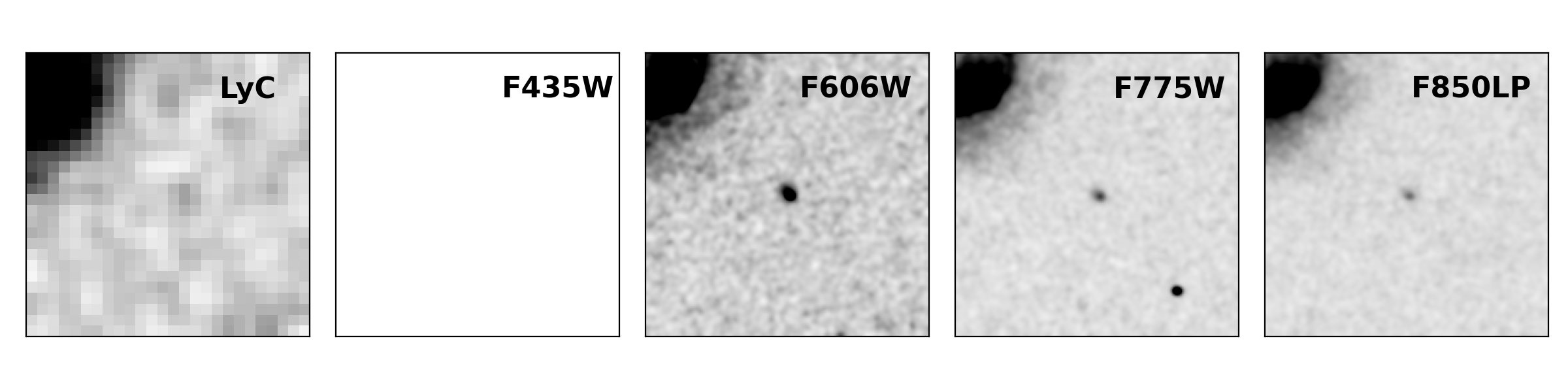}
    
    \vspace{-8pt}
    \includegraphics[width=0.48\textwidth]{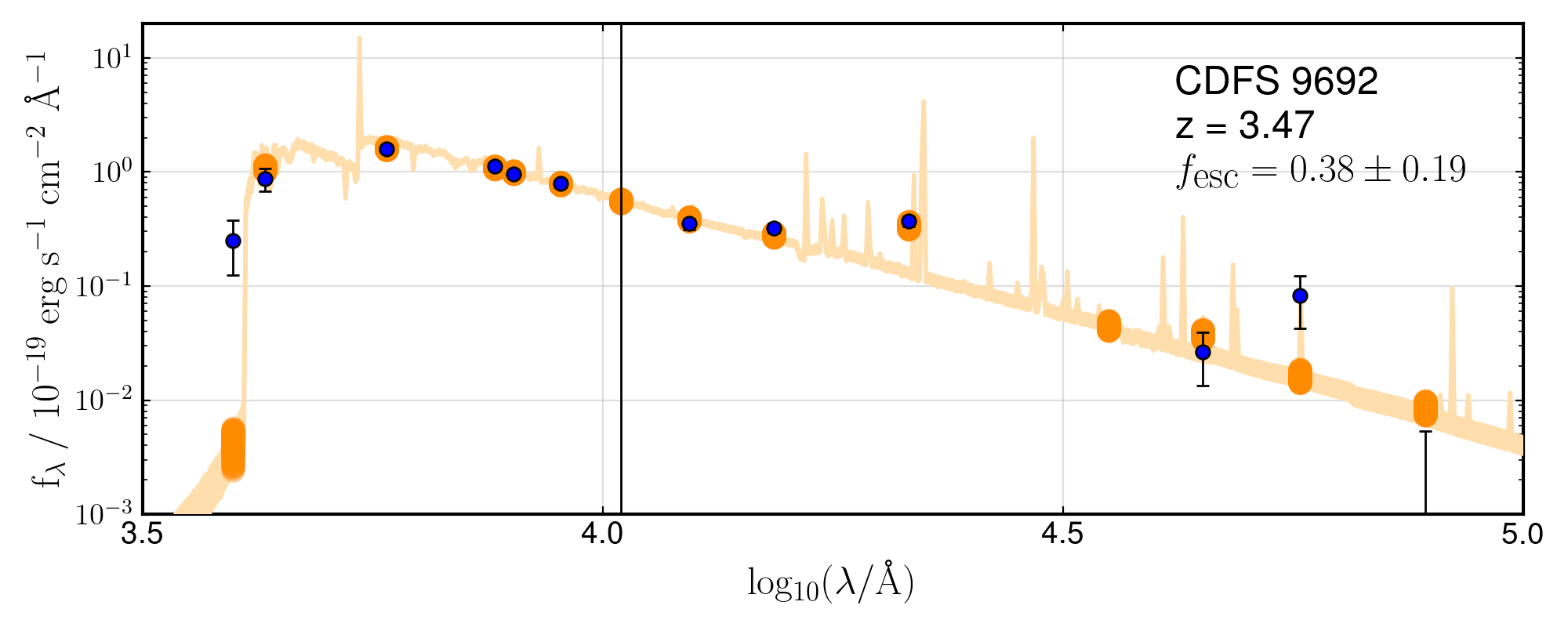}

    \includegraphics[width=0.48\textwidth]{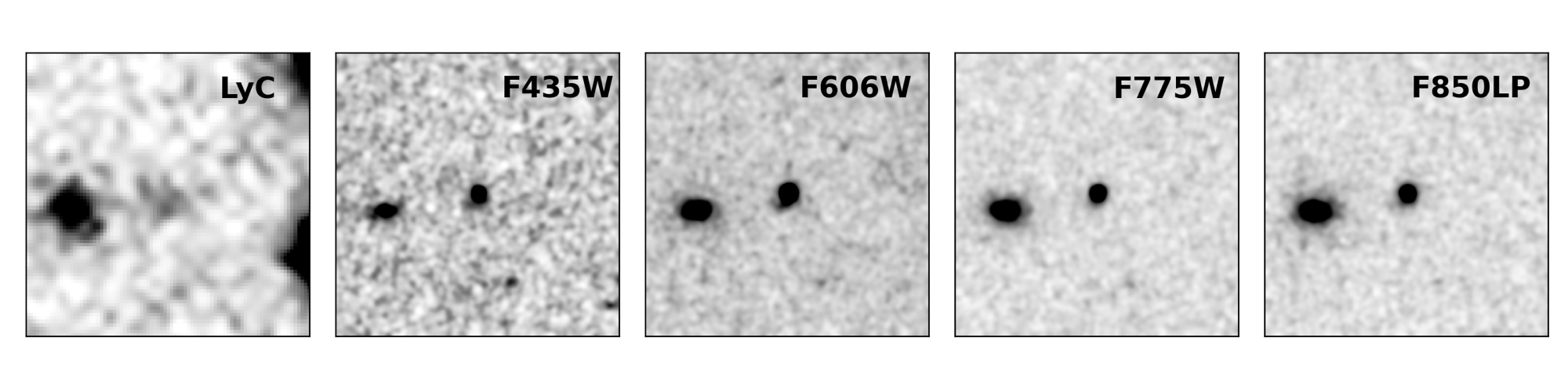}
    
    \vspace{-8pt}
    \includegraphics[width=0.48\textwidth]{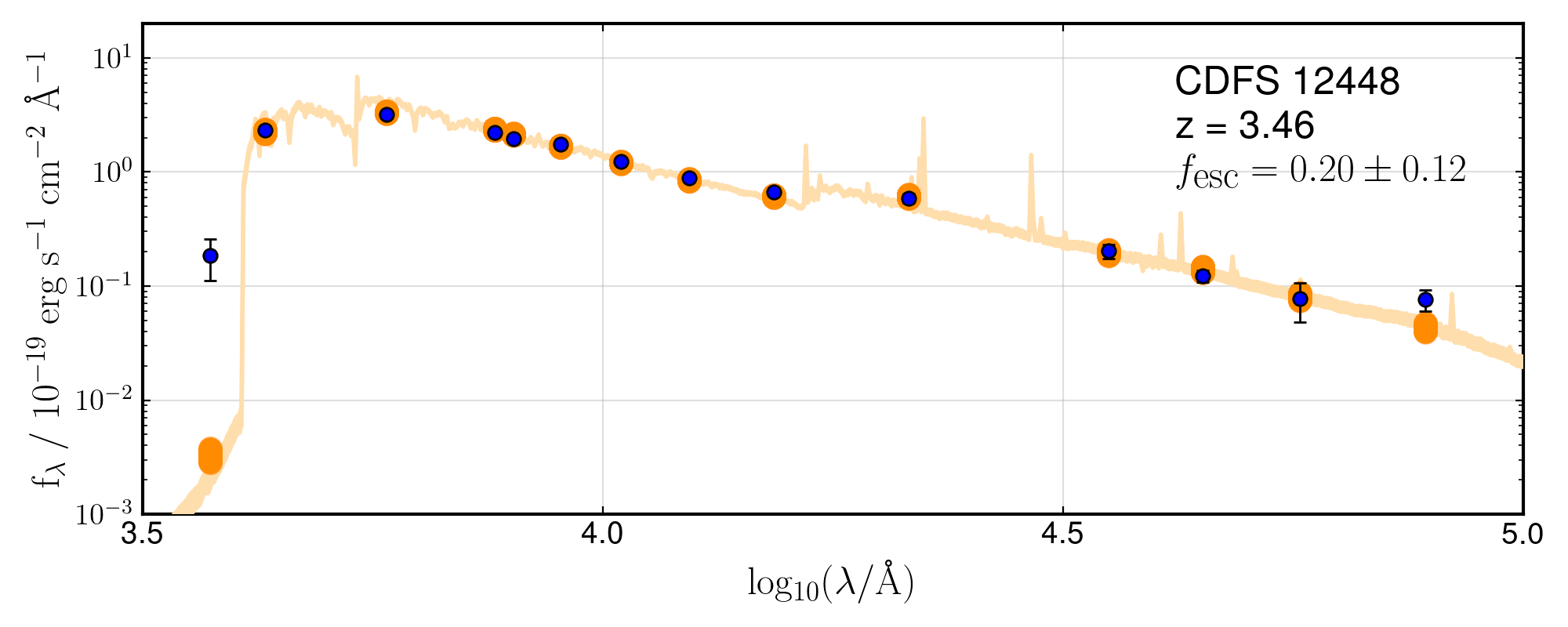}
   \caption{LyC and HST photometry (top), and best-fitting SED (orange) with observed (blue points) and SED derived (orange ovals) for candidate LyC leaking galaxies. We show the observed and SED predicted flux at LyC wavelengths only for demonstration purposes -- the SED derived LyC fluxes were not used to infer \fesc\, and the observed LyC fluxes were not used during SED fitting to obtain galaxy physical properties.}    
\end{figure}

\begin{figure}
\ContinuedFloat
    \centering
    \includegraphics[width=0.48\textwidth]{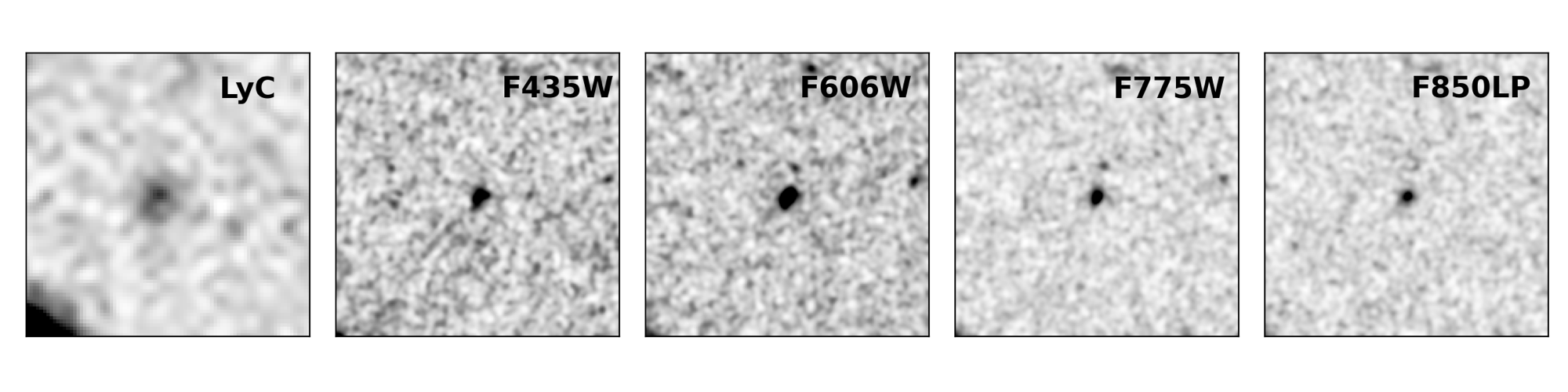}
    
    \vspace{-8pt}
    \includegraphics[width=0.48\textwidth]{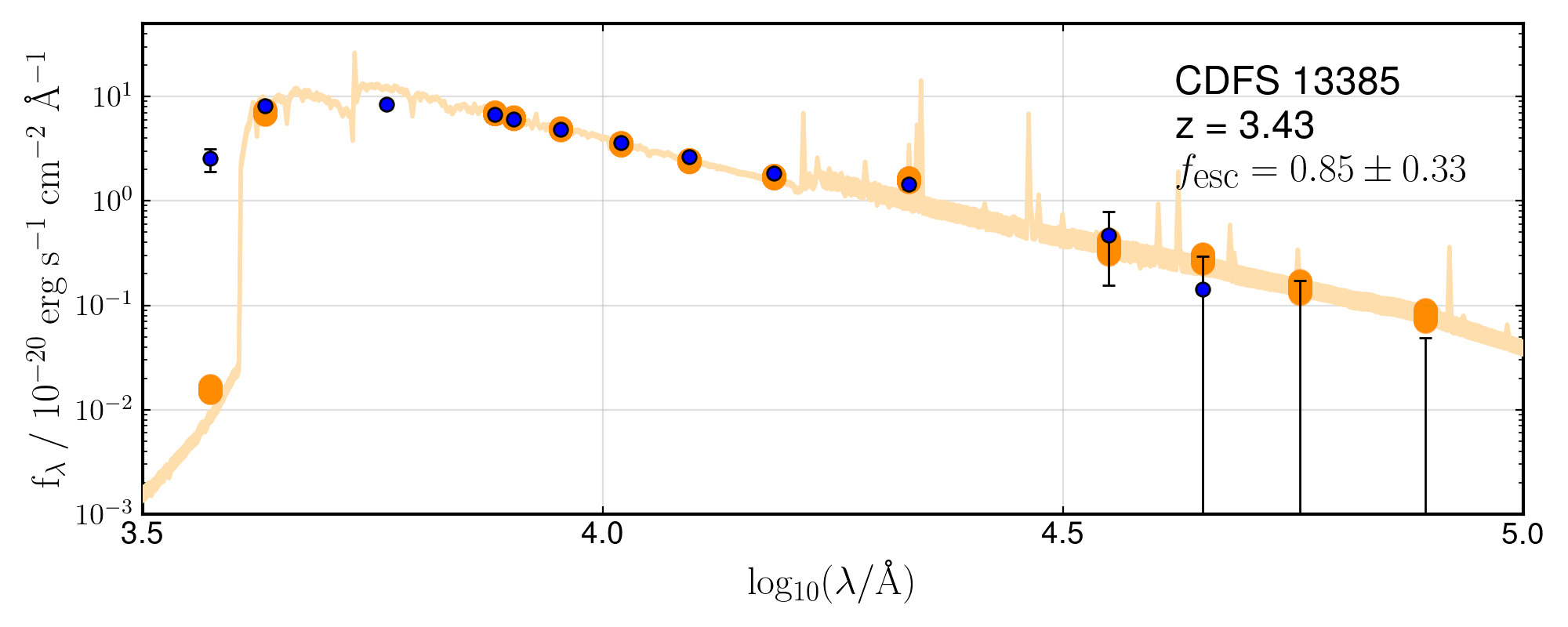}

    \includegraphics[width=0.48\textwidth]{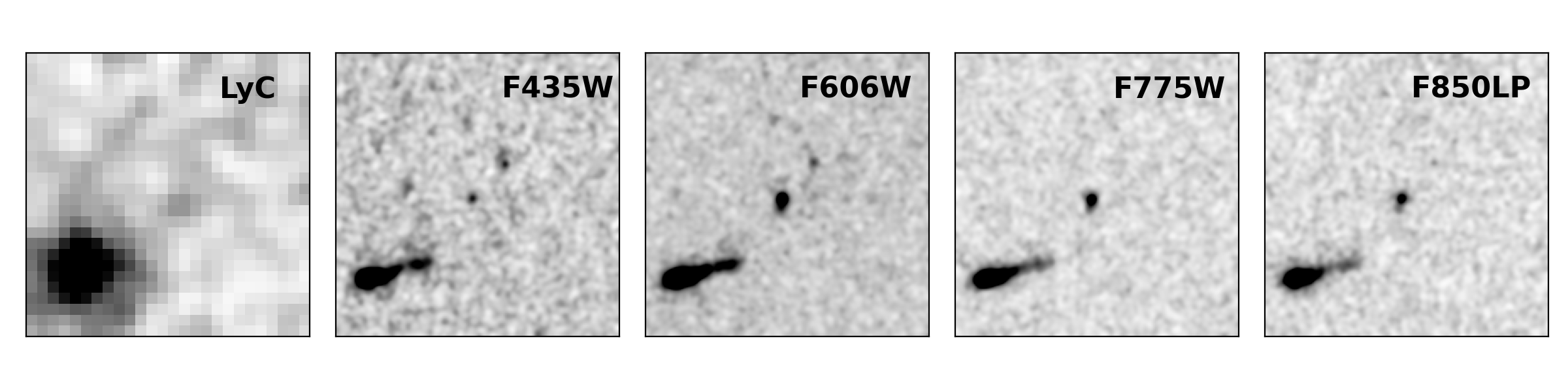}

    \vspace{-8pt}
    \includegraphics[width=0.48\textwidth]{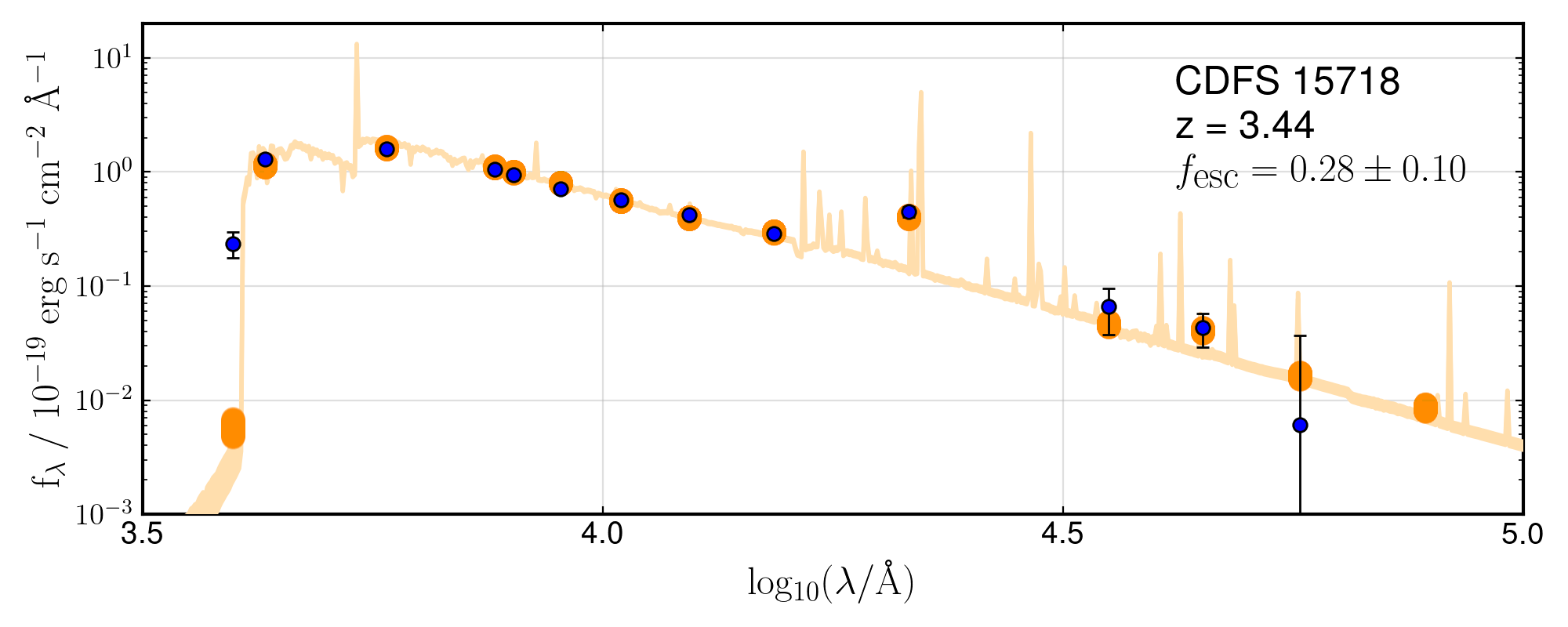}

    \includegraphics[width=0.48\textwidth]{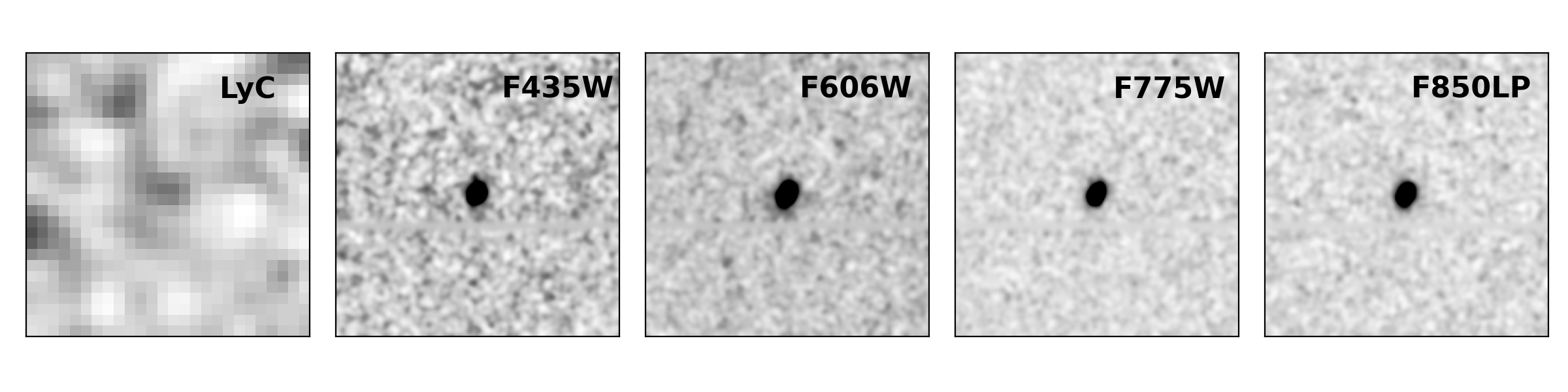}
    
    \vspace{-8pt}
    \includegraphics[width=0.48\textwidth]{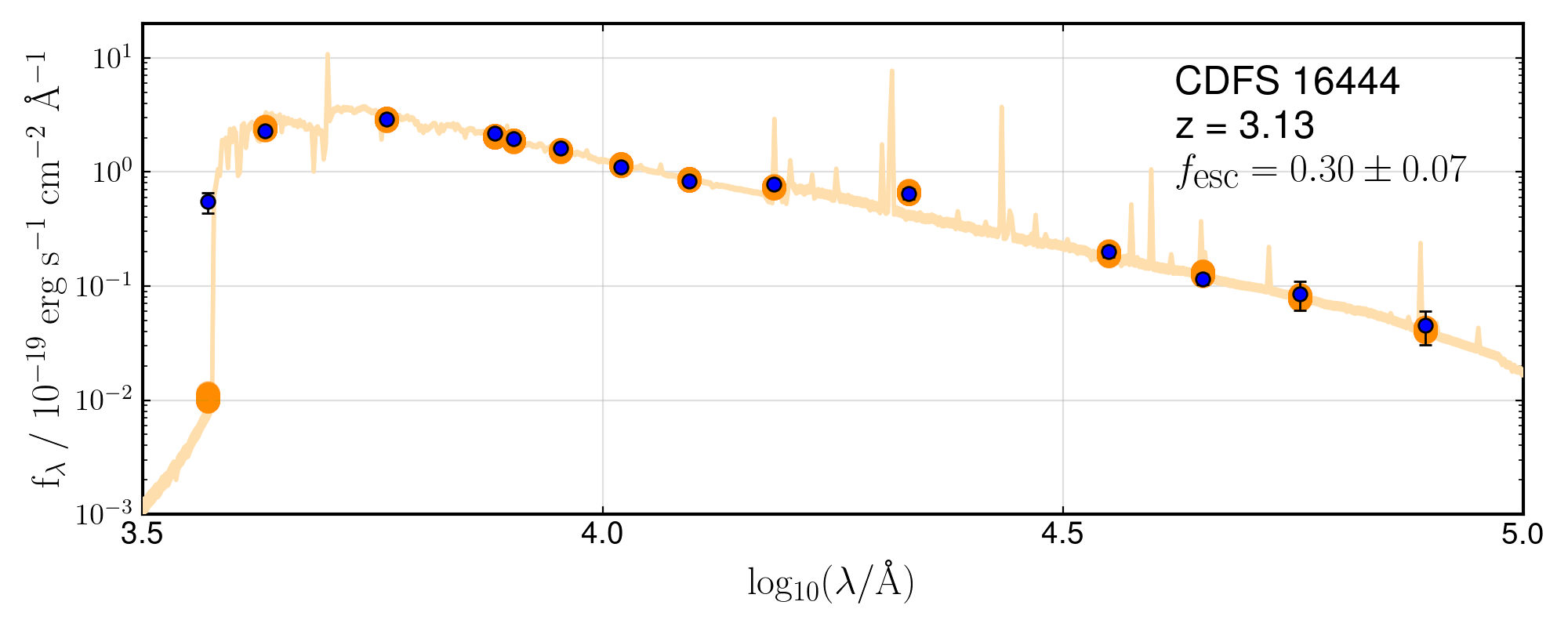}

    \includegraphics[width=0.48\textwidth]{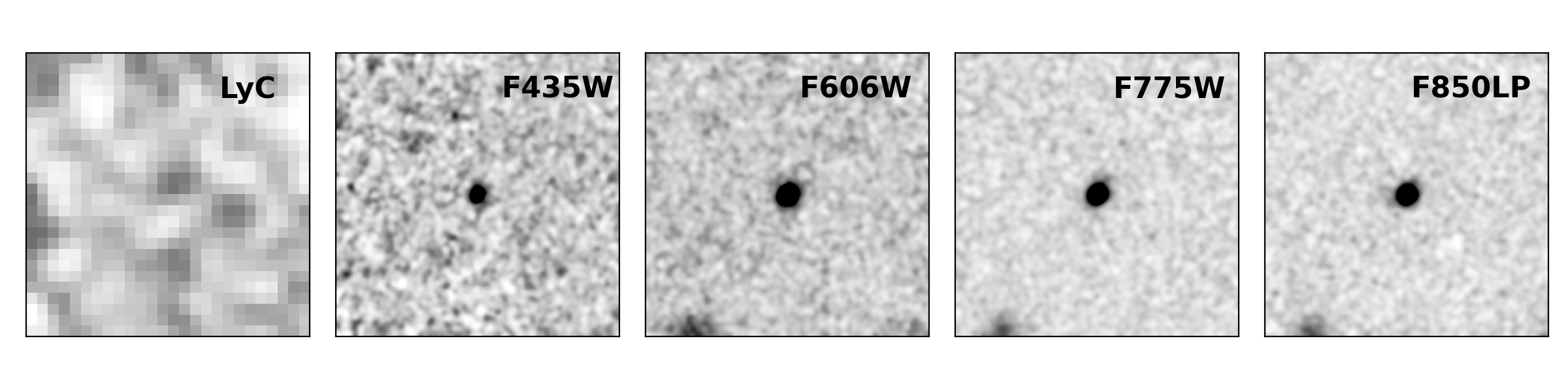}
    
    \vspace{-8pt}
    \includegraphics[width=0.48\textwidth]{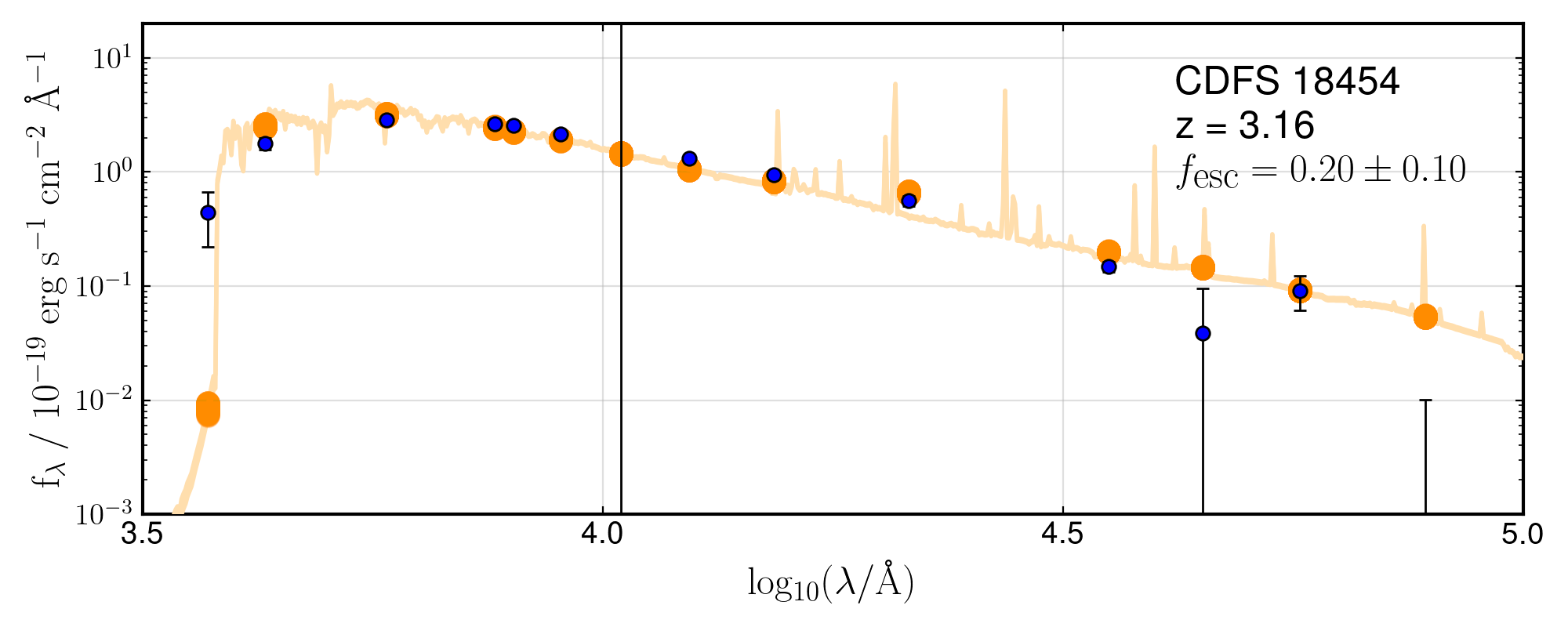}
    \caption{\textit{Continued}.}    
\end{figure}

\begin{figure}
\ContinuedFloat
    \centering
    \includegraphics[width=0.48\textwidth]{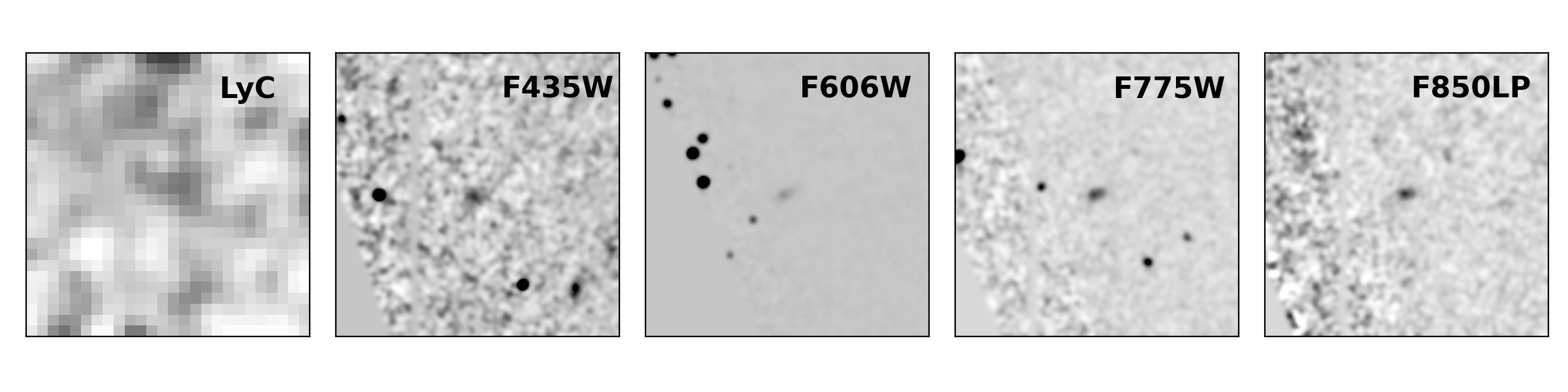}
    
    \vspace{-8pt}
    \includegraphics[width=0.48\textwidth]{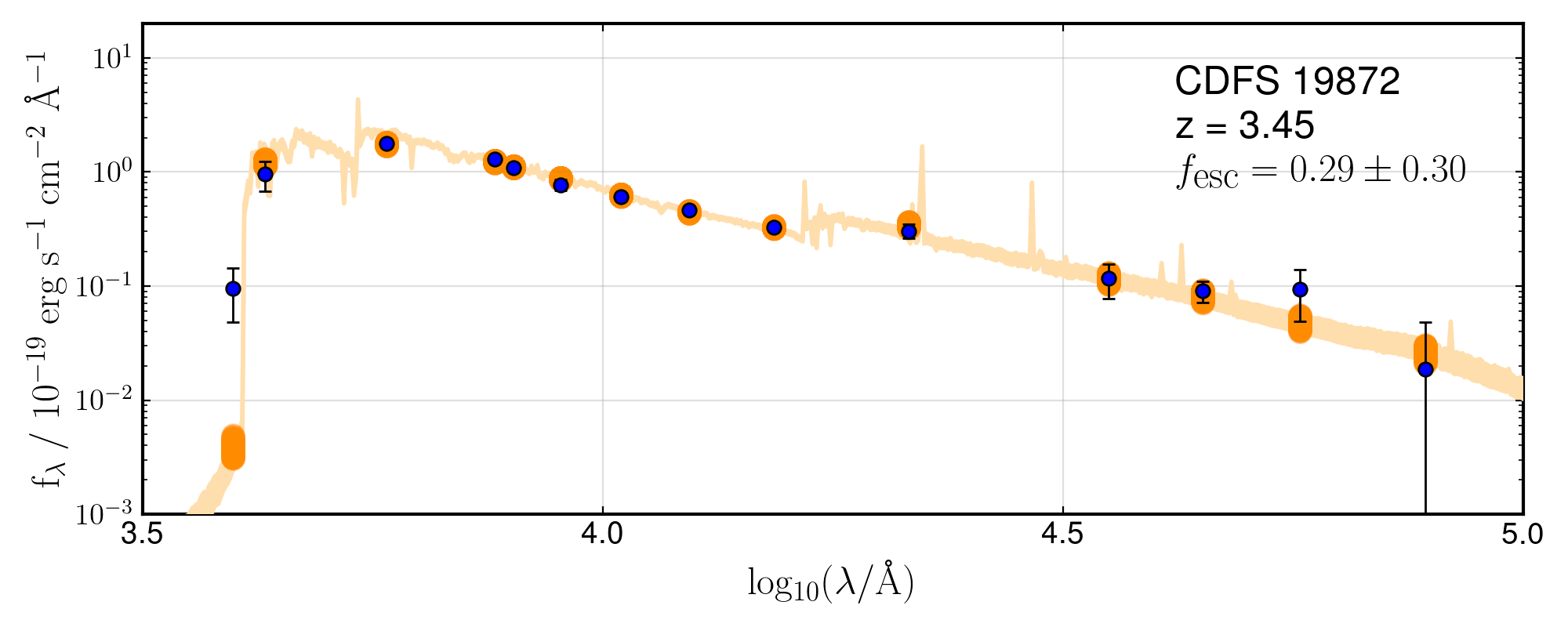}

    \includegraphics[width=0.48\textwidth]{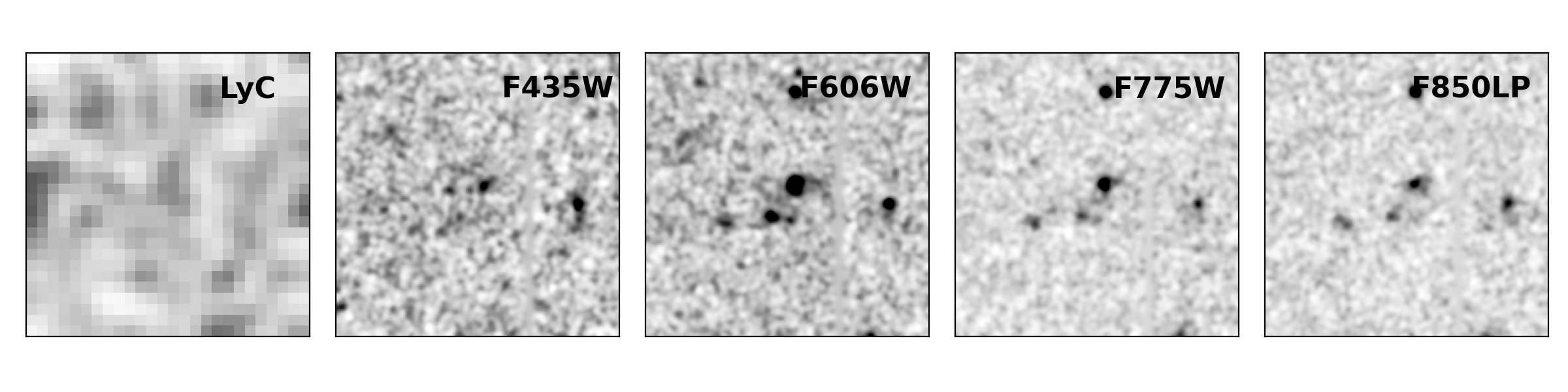}
    
    \vspace{-8pt}
    \includegraphics[width=0.48\textwidth]{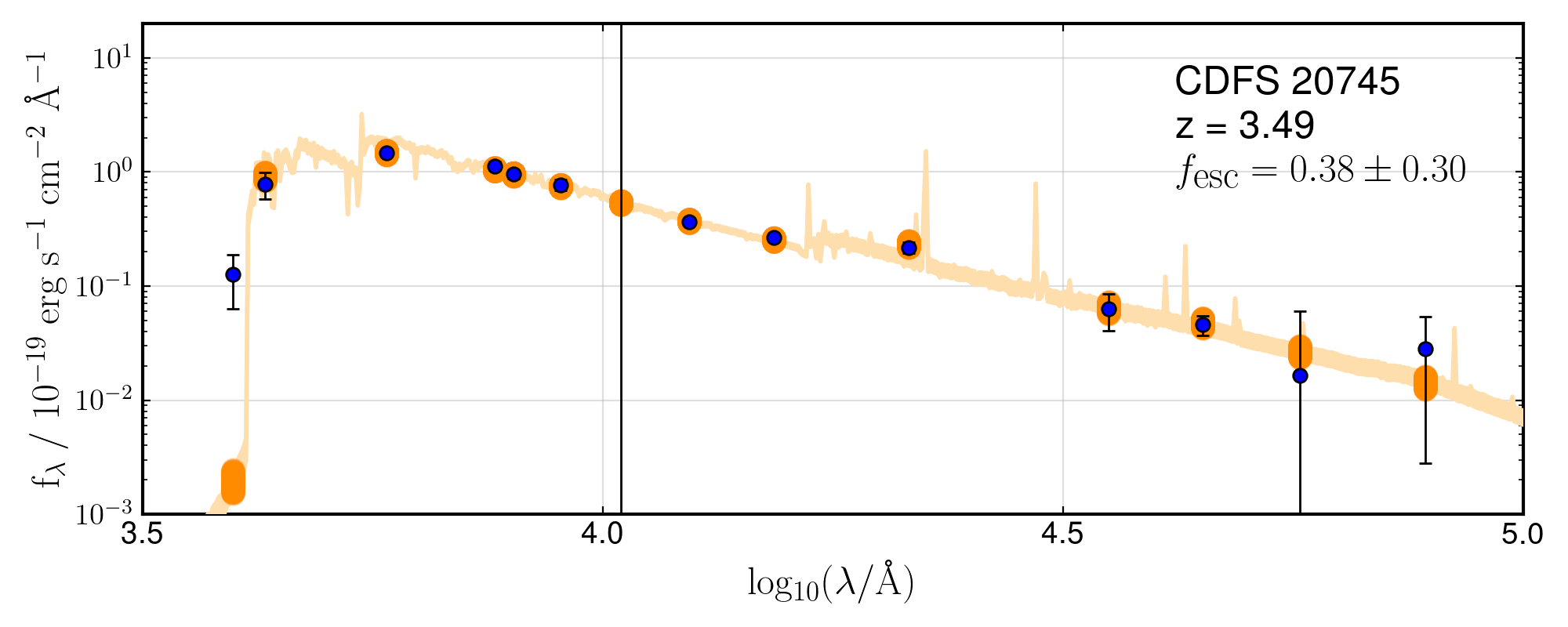}

    \includegraphics[width=0.48\textwidth]{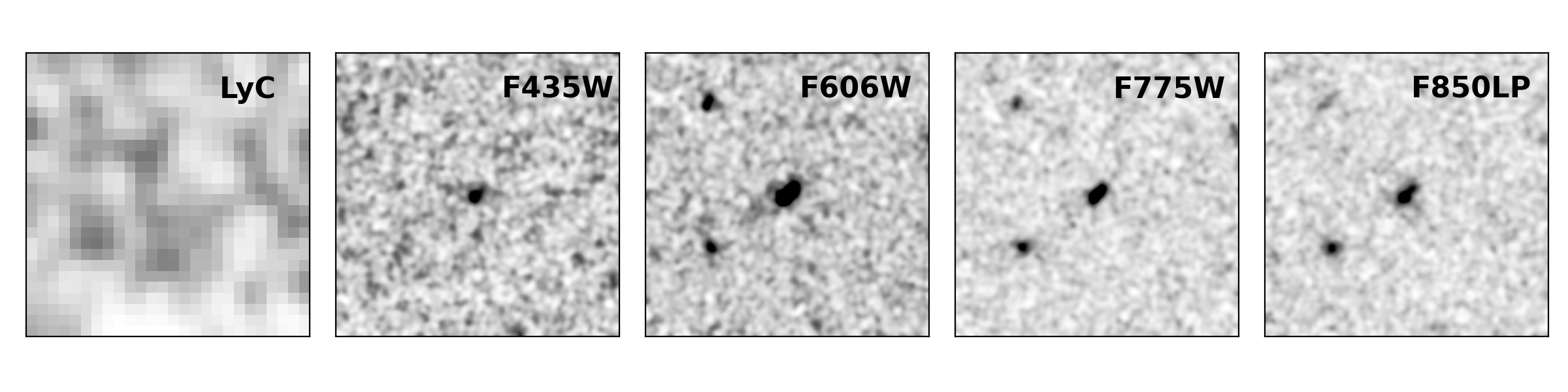}

    \vspace{-8pt}
    \includegraphics[width=0.48\textwidth]{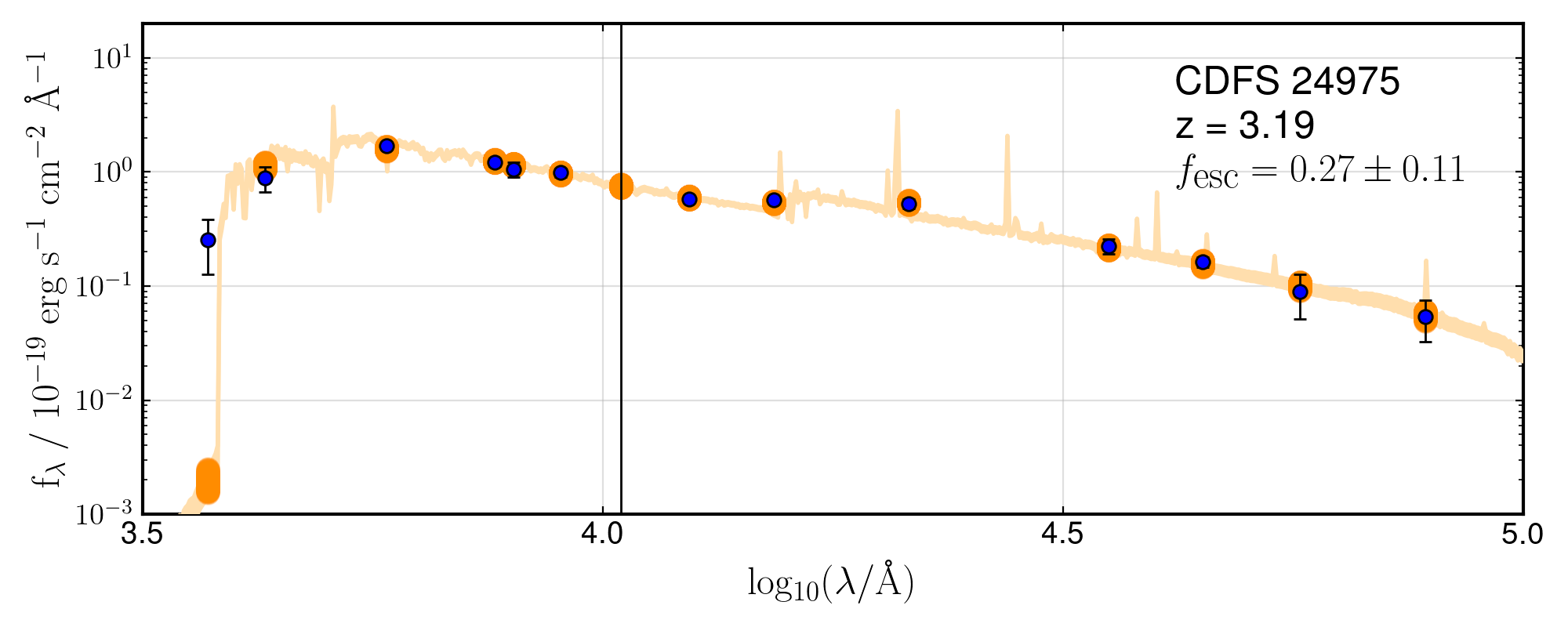}
    \caption{\textit{Continued}.}   
\end{figure}

\bsp	
\label{lastpage}
\end{document}